\documentclass[english,preprint,tightenlines,eqsecnum,aps,prd,nofootinbib]{revtex4-1}
\usepackage[T1]{fontenc}
\usepackage[latin9]{inputenc}
\usepackage{babel}
\usepackage{float}
\usepackage{amsmath}
\usepackage{amssymb}
\usepackage{graphicx}
\usepackage{esint}
\usepackage[unicode=true,pdfusetitle,
 bookmarks=true,bookmarksnumbered=false,bookmarksopen=false,
 breaklinks=false,pdfborder={0 0 1},backref=false,colorlinks=false]
 {hyperref}

\makeatletter

\let\SF@@footnote\footnote
\def\footnote{\ifx\protect\@typeset@protect
    \expandafter\SF@@footnote
  \else
    \expandafter\SF@gobble@opt
  \fi
}
\expandafter\def\csname SF@gobble@opt \endcsname{\@ifnextchar[
  \SF@gobble@twobracket
  \@gobble
}
\edef\SF@gobble@opt{\noexpand\protect
  \expandafter\noexpand\csname SF@gobble@opt \endcsname}
\def\SF@gobble@twobracket[#1]#2{}

\@ifundefined{textcolor}{}
{%
 \definecolor{BLACK}{gray}{0}
 \definecolor{WHITE}{gray}{1}
 \definecolor{RED}{rgb}{1,0,0}
 \definecolor{GREEN}{rgb}{0,1,0}
 \definecolor{BLUE}{rgb}{0,0,1}
 \definecolor{CYAN}{cmyk}{1,0,0,0}
 \definecolor{MAGENTA}{cmyk}{0,1,0,0}
 \definecolor{YELLOW}{cmyk}{0,0,1,0}
}

\usepackage{braket}
\usepackage{caption}
\@ifundefined{showcaptionsetup}{}{%
 \PassOptionsToPackage{caption=false}{subfig}}
\usepackage{subfig}
\makeatother

\begin{document}

\title{Pragmatic mode-sum regularization method for semiclassical black-hole
spacetimes }

\author{Adam Levi and Amos Ori}

\address{Department of physics, Technion-Israel Institute of Technology,\\
Haifa 3200, Israel }
\begin{abstract}
Computation of the renormalized stress-energy tensor is the most serious
obstacle in studying the dynamical, self-consistent, semiclassical
evaporation of a black hole in 4D. The difficulty arises from the
delicate regularization procedure for the stress-energy tensor, combined
with the fact that in practice the modes of the field need be computed
numerically. We have developed a new method for numerical implementation
of the point-splitting regularization in 4D, applicable to the renormalized
stress-energy tensor as well as to $\left\langle \phi^{2}\right\rangle _{ren}$,
namely the renormalized $\left\langle \phi^{2}\right\rangle $. So
far we have formulated two variants of this method: \emph{t-splitting}
(aimed for stationary backgrounds) and \emph{angular splitting} (for
spherically-symmetric backgrounds). In this paper we introduce our
basic approach, and then focus on the t-splitting variant, which is
the simplest of the two (deferring the angular-splitting variant to
a forthcoming paper). We then use this variant, as a first stage,
to calculate $\left\langle \phi^{2}\right\rangle _{ren}$ in Schwarzschild
spacetime, for a massless scalar field in the Boulware state. We compare
our results to previous ones, obtained by a different method, and
find full agreement. We discuss how this approach can be applied (using
the angular-splitting variant) to analyze the dynamical self-consistent
evaporation of black holes.
\end{abstract}
\maketitle

\section{Introduction}

After the discovery of Hawking radiation \cite{Hawking - Particle creation by black holes}
in 1975, it was widely anticipated that the semi-classical approach
to gravity based on quantum field theory (QFT) in curved spacetime
will open the opportunity to explore various interesting physical
problems, in which a quantum field interacts with the curved spacetime
where it resides. Among these problems,  of special interest is the
self-consistent dynamical evaporation of a black hole (BH). Another
outstanding problem in this class is the evolution of quantum fluctuations
in the very early universe, and the resulting cosmological structure
formation.

Indeed, in the latter problem of cosmological quantum-field perturbations
a remarkable progress has been achieved over the last few decades.
\cite{Parker - book,Parker - cosmology} However, the problem of analyzing
the semiclassical evolution of an evaporating BH still remains a serious
challenge. 

To understand the difficulties in analyzing this interesting problem
of self-consistent BH evaporation, let us briefly review the basic
structure of semiclassical gravity. The metric $g_{\alpha\beta}(x)$
is treated as a classical field, and it is assumed to satisfy the
\emph{semiclassical Einstein equation} 
\begin{equation}
G_{\alpha\beta}=8\pi\left\langle T_{\alpha\beta}\right\rangle _{ren},\label{eq: Intro - QFT-CST basic Eqs}
\end{equation}
where $G_{\alpha\beta}$ is the Einstein tensor associated with $g_{\alpha\beta}(x)$.
\footnote{Throughout this paper we use relativistic units $c=G=1$ and the $\left(-+++\right)$
signature.%
} The source term $\left\langle T_{\alpha\beta}\right\rangle _{ren}$
is the regularized expectation value of the stress-energy tensor associated
with a quantum field $\phi(x)$. For the sake of simplicity we shall
take $\phi$ here to be a scalar field. It is supposed to satisfy
the field equation 
\begin{equation}
\left(\square-m^{2}-\xi R\right)\phi=0,\label{eq:Field_Equation}
\end{equation}
where $m$ and $\xi$ respectively denote the mass and coupling constant
of the scalar field. Note that the theory is semi-classical, as the
field $\phi$ is quantized but the metric $g_{\alpha\beta}$ is classical,
nevertheless the two are coupled. 

The major difficulty in solving (or even analyzing) the field equation
(\ref{eq: Intro - QFT-CST basic Eqs}) has to do with the regularization
of the divergent quantity $\left\langle T_{\alpha\beta}\right\rangle $.
\footnote{There are also other difficulties, e.g. the runaway problem, as noted
by Wald \cite{Wald- QFT in CST}.%
} As opposed to QFT in flat spacetime in which one can use the normal-ordering
procedure, in curved spacetime the outcome of this procedure depends
on the choice of time slicing which is completely arbitrary. There
exists a regularization method named \emph{point-splitting} (PS),
also known as covariant point separation, which gives a general prescription
how to regularize quantities which are quadratic in the field and
its derivatives such as $\left\langle T_{\alpha\beta}\right\rangle $.
Alas, implementing this prescription in situations where the solution
of the field equation (\ref{eq:Field_Equation}) is known only numerically
turns out as a surprisingly difficult problem.

In this paper we shall focus on the regularization of $\left\langle \phi^{2}\right\rangle $
instead of $\left\langle T_{\alpha\beta}\right\rangle $. The quantity
$\left\langle \phi^{2}\right\rangle $ is also divergent, but not
as strong as $\left\langle T_{\alpha\beta}\right\rangle $. In addition,
the scalar character of $\left\langle \phi^{2}\right\rangle $ (as
opposed to the tensorial $\left\langle T_{\alpha\beta}\right\rangle $)
makes it easier to regularize. These properties make $\left\langle \phi^{2}\right\rangle $
a convenient tool to  examine and explain new ideas concerning regularization.
In order to calculate $\left\langle \phi^{2}\left(x\right)\right\rangle _{ren}$
(i.e. the regularized $\left\langle \phi^{2}\left(x\right)\right\rangle $)
using the PS method we split the point $x$ and write it as a product
of $\phi$ at two different points, namely $\left\langle \phi\left(x\right)\phi\left(x'\right)\right\rangle $.
This is known as the \emph{two-point function} (TPF). We then subtract
from the TPF a known \textit{counter-term} and take the limit $x'\to x$.
This limit, however, is what makes the numerical implementation so
hard.

In 1984 Candelas and Howard \cite{Candelas =000026 Howard - 1984 - phi2 Schwrazschild}
developed a method to numerically implement PS if a high-order WKB
approximation for the field modes is known. Using this method, they
calculated $\left\langle \phi^{2}\right\rangle _{ren}$ in Schwarzschild
spacetime, and subsequently this method was used by Howard \cite{Howard - 1984 - Tab Schwarzschild}
to calculate $\left\langle T_{\alpha\beta}\right\rangle _{ren}$ in
Schwarzschild. Later, this method was extended to a general static
spherically symmetric background, first for $\left\langle \phi^{2}\right\rangle _{ren}$
by Anderson \cite{Anderson - 1990 - phi2 static spherically symmetric},
and subsequently to $\left\langle T_{\alpha\beta}\right\rangle _{ren}$
by Anderson, Hiscock and Samuel \cite{Anderson - 1995 - Tab static spherically symmetric}.

The limitation of the method proposed by Candelas and Howard is that
it requires a high-order WKB approximation for the field modes: at
least second-order for $\left\langle \phi^{2}\right\rangle _{ren}$
and forth order for $\left\langle T_{\alpha\beta}\right\rangle _{ren}$.
In the ordinary (Lorentzian-signature) Schwarzschild metric this task
of high-order WKB expansion is very difficult, especially because
of the presence of a turning point: For typical modes of large $\omega$
and $l$, there is a turning point on the $r$ axis, at a value $r_{turn}(\omega,l)$
where the effective potential $V_{l}(r)$ equals $\omega^{2}$. %
\footnote{In Schwarzschild spacetime (and any other eternal BH spacetimes),
for given $\omega$ and $l$ there are usually \emph{two} such roots.
Here we shall explicitly refer to the larger one, but the same complications
arise also at the smaller root. %
} The mode's radial function is essentially oscillatory at $r>r_{turn}$
and exponential at $r<r_{turn}$. Both these basic WKB approximations
--- the oscillatory approximation at $r>r_{turn}$ and the exponential
approximation at $r<r_{turn}$ --- break down and actually diverge
at $r\rightarrow r_{turn}$. To correctly match the two approximations,
one has to use another, intermediate approximation valid in the neighborhood
of $r=r_{turn}$. This \emph{turning-point approximation} is based
on the Airy function. Whereas the leading-order matching is manageable,
it becomes exceedingly hard to go to higher-order WKB, because each
succeeding order will now require its own turning-point matching.
Furthermore, to the best of our understanding, the series of powers
involved  in the Airy-based turning-point expansion proceeds in powers
of $\omega^{-1/3}$ (rather than $\omega^{-1}$). Correspondingly,
we may expect that in the presence of a turning point, to implement
the WKB-based expansion to order $\omega^{-4}$ (required for calculating
$\left\langle T_{\alpha\beta}\right\rangle _{ren}$), one would have
to carry matched asymptotic expansion up to \emph{twelfth} order in
$\omega^{-1/3}$ (or six such orders for $\left\langle \phi^{2}\right\rangle _{ren}$)
--- a formidably difficult task.

To overcome these difficulties, Candelas and Howard \cite{Candelas =000026 Howard - 1984 - phi2 Schwrazschild}
and several others \cite{Howard - 1984 - Tab Schwarzschild,Anderson - 1990 - phi2 static spherically symmetric,Anderson - 1995 - Tab static spherically symmetric}
used an elegant trick: They used Wick rotation to analytically extend
the background metric to the Euclidean sector. Any static spacetime
is guaranteed to have such a real-metric Euclidean sector. In the
latter, the radial equation does not admit a turning point. This way,
it was possible to carry the WKB analysis to the desired order and
to implement the above regularization scheme --- for $\left\langle \phi^{2}\right\rangle _{ren}$
as well as for $\left\langle T_{\alpha\beta}\right\rangle _{ren}$. 

Our ultimate goal, however, is to develop a regularization scheme
for $\left\langle \phi^{2}\right\rangle $ and $\left\langle T_{\alpha\beta}\right\rangle $,
applicable to \emph{time-dependent} backgrounds as well. Such a time-dependent
metric (even if spherically symmetric) does $not$ generically admit
a Euclidean sector. We are thus led to carry the analysis directly
in the Lorentzian sector, which in turn implies the presence of a
turning point in the radial equation, hampering any attempt to carry
high-order WKB expansion. We shall therefore refrain from establishing
our regularization scheme on the WKB analysis. 

There is another obvious reason for avoiding WKB analysis: Consider
a time-dependent spherically-symmetric background. The field equation
for a given mode may still be expressed as a one-dimensional (namely
$1+1$) wave equation with an effective potential, but now the potential
will be time-dependent. In such a situation, even the leading-order
WKB (and even if we forget for the moment about the turning point)
becomes a non-trivial task, let alone higher-order WKB analysis.

For these reasons, we shall not base our regularization scheme on
high-order WKB expansion. Instead, in our method we extract the required
information concerning the high-frequency field's modes directly from
the well-known counter-term (\ref{eq: basic PS - The counter-terms GDS})
for $\left\langle \phi^{2}\right\rangle _{ren}$ (and, for computing
$\left\langle T_{\alpha\beta}\right\rangle _{ren}$, from the counter-term
developed by Christensen \cite{Christiansen}). 

We point out that this approach, namely extraction of the high-frequency
asymptotic behavior of the modes from a known local counter-term,
was actually initiated by Candelas \cite{Candelas - 1979 - phi2 Schwarzschild}
--- already before he and Howard resorted to the Euclidean sector
\cite{Candelas =000026 Howard - 1984 - phi2 Schwrazschild}. However,
Candelas' analysis was restricted to the Schwarzschild case (which
in particular means restriction to staticity, spherical symmetry,
and to vacuum %
\footnote{In particular, the logarithmic counter-term is not encountered in
the vacuum case.%
}). We should also comment that even in that case the analysis in Ref.
\cite{Candelas - 1979 - phi2 Schwarzschild} was not completed, because
the required integral of the regularized mode contribution over $\omega$
was not carried out. When we attempted to implement this integral
over $\omega$, we found that it actually fails to converge in the
usual sense, due to growing oscillations (see below), which led us
to introduce the notion of generalized integral. Our approach is in
this sense a completion of Candelas' method, as well as its generalization
beyond the vacuum case (and with the scope of further extending it
to dynamical backgrounds). 

Our method requires the field modes to admit a trivial decomposition
in at least one of the coordinates (e.g. through $e^{-i\omega t}$
or spherical harmonics). This usually corresponds to having a Killing
field in spacetime. %
\footnote{Having at least one trivial coordinate is a necessary condition for
the applicability of our method, but we do not claim that it is also
a sufficient condition. %
} The splitting is then done in that trivial coordinate --- which enables
us to treat the coincidence limit analytically. So far we have developed
two different variants of our method: (i) the \emph{$t$-splitting}
variant, which requires a time-translation symmetry; (ii) the \emph{angular-splitting}
variant, which requires spherical symmetry. We are also exploring
a third variant, \emph{azimuthal splitting} (which would only require
axial symmetry), but this one is still in progress. In all these variants
we assume for simplicity that the background is asymptotically-flat,
although this requirement can probably be relaxed.

Since our ultimate goal is to analyze $\left\langle T_{\alpha\beta}\right\rangle _{ren}$
on the time-dependent background of an evaporating BH, the $t$-splitting
method is insufficient, and we shall actually need the angular-splitting
variant. It turns out, however, that the $t$-splitting variant is
in some sense conceptually simpler and easier to present at first
stage, because certain additional complications arise in the angular-splitting
method. Although we know how to address these complications, they
make the method's logical structure a bit more obscure and harder
to explain. For this reason, for the sake of introducing our basic
regularization strategy we choose to present here the $t$-splitting
variant, which is logically simpler. And for exactly the same reason,
in this paper we shall display the regularization of $\left\langle \phi^{2}\right\rangle $
rather than $\left\langle T_{\alpha\beta}\right\rangle $. Then in
the next paper we plan to present the angular-splitting variant, which
is going to be our main tool for analyzing dynamical BH evaporation. 

The TPF diverges for any pair of points connected by a null geodesic,
even if they are far from each other \cite{Kay Radzikowski and Wald - 1996}.
As it turns out, this long-distance divergence of the TPF leads to
undamped oscillations in the mode contributions at large $\omega$.
To address this issue we use the concept of \emph{generalized integral},
in which these oscillations are properly damped upon integration over
$\omega$, which fully cures the oscillations problem. The origin
of this complication (the presence of connecting null geodesics) is
discussed in Sec. \ref{sub: Remarks-about-the-TPF} and also in Appendix
\ref{sec: Appendix-B}; And the resolution of the oscillations problem
by means of generalized integral (and particularly the so called ``self-cancellation
integral'') is described in Sec. \ref{sub: Remarks-about-the-TPF}
and further in Appendix \ref{sec: Appendix-A}. 

In the description of the $t$-splitting method in Sec. \ref{sec:The-new-scheme}
we assume a spherically-symmetric static background for the sake of
simplicity. However, as was discussed above, the $t$-splitting method
does not require spherical symmetry, and in principle it may be applied
to a generic (asymptotically-flat) stationary spacetime. We outline
this generalization of the method to stationary backgrounds in Sec.
\ref{sub: Our method - Non spherical symmetric}. We point out, however,
that some completion is still required in the case of a stationary
eternal BH (see therein). 

Next we apply our method explicitly to the Schwarzschild case, computing
$\left\langle \phi^{2}\right\rangle _{ren}$ in Boulware state. We
compare our results to those obtained previously by Anderson (using
the Euclidean sector), and find full agreement. 

This article is divided as follows: In Sec. \ref{sec: Basic point splitting}
we present the basic PS method, and then briefly outline the procedure
developed by Candelas and Howard \cite{Candelas =000026 Howard - 1984 - phi2 Schwrazschild}.
Note that Sec. \ref{sub: Remarks-about-the-TPF} discusses certain
subtleties of the TPF in some detail, and may be skipped in first
reading. In Sec. \ref{sec:The-new-scheme} we present our $t$-splitting
method. We first describe it for spherically-symmetric static background,
and then outline its generalization to a generic stationary background.
In Sec. \ref{sec: Calc in Schwarzschild } we harness this method
for the calculation of $\left\langle \phi^{2}\right\rangle _{ren}$
in the Schwarzschild metric. Finally, in Sec. \ref{sec: Discussion}
we discuss the implications of our new method and try to pave the
path towards our ultimate goal of investigating self-consistent BH
evaporation.

\section{Basic point-splitting method and its numerical implementation \label{sec: Basic point splitting}}

We start by sketching the basic PS regularization method. This method
is aimed to regularize the expectation value of various quantities
which are quadratic in the field operator (and its derivatives). Among
these quantities, the most important one is probably the energy-momentum
tensor $\left\langle T_{\alpha\beta}\right\rangle $. However, in
this first paper we shall consider $\left\langle \phi^{2}\right\rangle $
as a simpler example (although we shall occasionally remark on the
analogous calculation of $\left\langle T_{\alpha\beta}\right\rangle $). 

For simplicity we consider here a quantum scalar field $\phi(x)$
living in a static, spherically symmetric, asymptotically flat spacetime
with metric
\begin{equation}
ds^{2}=g_{tt}(r)dt^{2}+g_{rr}(r)dr^{2}+r^{2}d\Omega^{2},\label{eq:  metric}
\end{equation}
where $d\Omega^{2}\equiv d\theta^{2}+\sin^{2}\theta\, d\varphi^{2}$.
The field operator may then conveniently be expressed as
\begin{equation}
\phi\left(x\right)=\int_{0}^{\infty}d\omega\sum_{l=0}^{\infty}\sum_{m=-l}^{l}\left(f_{\omega lm}\left(x\right)a_{\omega lm}+f_{\omega lm}^{*}\left(x\right)a_{\omega lm}^{\dagger}\right).\label{eq: basic PS - field decomposition to a,ad}
\end{equation}
Here, $a_{\omega lm}^{\dagger}$ and $a_{\omega lm}$ are the creation
and annihilation operators of the field's $\omega lm$ mode, $f_{\omega lm}\left(x\right)$
is a complete, orthonormal, family of modes %
\footnote{By ``orthonormal'' we mean that the inner product of two mode functions
$f_{\omega lm}$ and $f_{\omega'l'm'}$ is $\delta_{ll'}\delta_{mm'}\delta(\omega-\omega')$.%
} taking the form 

\begin{equation}
f_{\omega lm}\left(x\right)=e^{-i\omega t}Y_{lm}\left(\theta,\varphi\right)\bar{\psi}_{\omega l}\left(r\right),\label{eq: basic PS - expression for f}
\end{equation}
and $Y_{lm}\left(\theta,\varphi\right)$ are the usual spherical harmonics.
The radial functions $\bar{\psi}_{\omega l}\left(r\right)$ are obtained
by solving the field equation for $\phi(x)$ with the decomposition
(\ref{eq: basic PS - expression for f}) and with appropriate boundary
conditions. 

The operation of summation over $m,l$ and integration over $\omega$
repeats many times in the analysis below. We shall generally refer
to this operation as the ``mode sum'' (despite the slight abuse
of terminology). We point out, however, that the decomposition (\ref{eq: basic PS - field decomposition to a,ad})
applies as-is in the case of an asymptotically-flat background spacetime
with simple asymptotic structure (like e.g. Minkowski or a star).
But if the background spacetime is an eternal BH with a past horizon,
then for each $\omega lm$ combination there are actually two orthonormal
modes, namely the ``in'' and ``up'' modes. The ``in'' modes
are those described above (namely monochromatic waves propagating
from past null infinity), and the ``up'' modes describe monochromatic
waves that emerge from the past horizon. In this case of eternal BH,
the mode sum should also include a summation over the contributions
of these two independent modes for each $\omega lm$ (as explicitly
described in Sec. \ref{sec: Calc in Schwarzschild } for the Schwarzschild
case). 

The quantity $\left\langle \phi^{2}\right\rangle $ obviously depends
on the quantum state. Naturally, one would like to evaluate it in
the vacuum state. The latter is defined to be the quantum state annihilated
by each of the above $a_{\omega lm}$ operators. %
\footnote{If a past horizon exists, then one need to further specify this vacuum
state, e.g. by prescribing the outcome of the action of the ``up''
annihilation operators on that state. In the analysis of the Schwarzschild
case in Sec. \ref{sec: Calc in Schwarzschild } we shall consider
the Boulware vacuum state.%
}

Trying to naively calculate $\left\langle \phi^{2}\left(x\right)\right\rangle $
in the vacuum state yields the divergent expression
\begin{equation}
\left\langle \phi^{2}\left(x\right)\right\rangle _{naive}=\hbar\int_{0}^{\infty}d\omega\sum_{l=0}^{\infty}\sum_{m=-l}^{l}\left|Y_{lm}\left(\theta,\varphi\right)\right|^{2}\left|\bar{\psi}_{\omega l}\left(r\right)\right|^{2}.\label{eq: basic PS - phi2 naive mode sum}
\end{equation}
Although the sum over $m,l$ does converge for a given $\omega$,
the integral over $\omega$ diverges. In fact, one can easily check
that already in Minkowski spacetime the integrand is $\propto\omega$
(and $\propto\omega^{3}$ for $\left\langle T_{\alpha\beta}\right\rangle $),
and the same divergence occurs in the Schwarzschild case as well.
If one tries to integrate over $\omega$ before the summation, one
finds that the integral over $\omega$ again diverges (for a given
$l,m$). 

One therefore needs to somehow regularize the expression for $\left\langle \phi^{2}\right\rangle $.
In flat spacetime this regularization may be easily achieved by normal
ordering, but in curved spacetime this yields a slicing-dependent,
non-unique result. This is where point splitting comes into play.
DeWitt \cite{Dewitt - Dynamical theory of groups and fields} proposed
that $\left\langle \phi^{2}\left(x\right)\right\rangle $ (and, more
generally, quantities quadratic in the field operator and its derivatives)
can be treated by taking the product of the field operators in two
separate points $x,x'$ and then considering the coincidence limit
$x\rightarrow x'$. More specifically, he showed that the regularized
expectation value of $\phi^{2}$ can be defined as
\begin{equation}
\left\langle \phi^{2}\left(x\right)\right\rangle _{ren}=\lim_{x'\to x}\left[\left\langle \phi\left(x\right)\phi\left(x'\right)\right\rangle -G_{DS}\left(x,x'\right)\right].\label{eq: basic PS - phi Ren limit}
\end{equation}
Here $G_{DS}\left(x,x'\right)$ is the DeWitt-Schwinger \textit{counter-term},
namely a local term which fully captures the singular piece of the
TPF. For a scalar field with mass $m$ and coupling constant $\xi$
it takes the form \cite{Anderson - 1995 - Tab static spherically symmetric}
\begin{multline}
\frac{1}{\hbar}G_{DS}\left(x,x'\right)=\frac{1}{8\pi^{2}\sigma}+\frac{m^{2}+\left(\xi-1/6\right)R}{8\pi^{2}}\left[\gamma+\frac{1}{2}\ln\left(\frac{\mu^{2}\sigma}{2}\right)\right]\\
-\frac{m^{2}}{16\pi^{2}}+\frac{1}{96\pi^{2}}R_{\alpha\beta}\frac{\sigma^{;\alpha}\sigma^{;\beta}}{\sigma}.\label{eq: basic PS - The counter-terms GDS}
\end{multline}
Here $R$ and $R_{\alpha\beta}$ are respectively the Ricci scalar
and tensor, $\gamma$ denotes the Euler constant, and $\sigma\left(x,x'\right)$
is the biscalar associated with the short geodesic connecting $x$
and $x'$. The value of $\sigma$ is half the geodesic distance squared
(see Ref. \cite{Christiansen}). More specifically, for a timelike
separation $\sigma=-\tau^{2}/2$, where $\tau$ denotes the proper
time between $x$ and $x'$.  The parameter $\mu$ is unknown and
it corresponds to the well-known ambiguity in the regularization procedure
\cite{Wald- QFT in CST}. %
\footnote{In the case of a massive scalar field some authors took $\mu$ to
be the field's mass $m$. \cite{Christiansen,Anderson - 1995 - Tab static spherically symmetric}%
}

\subsection{Previous numerical implementations of point splitting}

It is not easy to directly implement the point-splitting method, especially
because in the cases of interest (e.g. black hole backgrounds) the
radial functions $\bar{\psi}_{\omega l}$ are only known from numerics.
Furthermore, in Eq. (\ref{eq: basic PS - phi Ren limit}) the TPF
$\left\langle \phi\left(x\right)\phi\left(x'\right)\right\rangle $,
which is to be computed numerically, diverges as $\tau^{-2}$ (and
as $\tau^{-4}$ for $\left\langle T_{\alpha\beta}\right\rangle $)
as $x'$ approaches $x$, where $\tau$ denotes the geodesic distance
between $x$ and $x'$. 

This problem was addressed by Candelas, Howard and later Anderson
and collaborators (see Refs. \cite{Candelas =000026 Howard - 1984 - phi2 Schwrazschild,Howard - 1984 - Tab Schwarzschild,Anderson - 1990 - phi2 static spherically symmetric,Anderson - 1995 - Tab static spherically symmetric})
a long time ago. They developed a calculation scheme that allows regularization
of $\left\langle \phi^{2}\right\rangle $ or $\left\langle T_{\alpha\beta}\right\rangle $
numerically, provided that an analytic approximation for the field
is known, up to a sufficiently high order %
\footnote{In general it would require a second order WKB approximation in order
to compute $<\phi^{2}>_{ren}$, and a fourth order approximation to
compute $<T_{\alpha\beta}>_{ren}$. %
}. 

In that scheme, one first analytically constructs the approximate
singular piece of the field, which we denote $\phi_{sing}(x)$. It
is composed of the contribution of the modes of large $\omega$ and
$l$, up to second order in $1/\omega$ and $1/l$ (fourth order for
$\left\langle T_{\alpha\beta}\right\rangle $), which is usually computed
using WKB analysis. From this quantity one then constructs the approximate
TPF $\left\langle \phi_{sing}(x)\phi_{sing}(x')\right\rangle $, and
recasts Eq. (\ref{eq: basic PS - phi Ren limit}) in the form 
\begin{multline}
\left\langle \phi^{2}\left(x\right)\right\rangle _{ren}=\lim_{x'\to x}\left[\left\langle \phi\left(x\right)\phi\left(x'\right)\right\rangle -\left\langle \phi_{sing}(x)\phi_{sing}(x')\right\rangle \right]\\
+\lim_{x'\to x}\left[\left\langle \phi_{sing}(x)\phi_{sing}(x')\right\rangle -G_{DS}\left(x,x'\right)\right].\label{eq: basic PS - phi Ren two limits}
\end{multline}
Since $\left\langle \phi_{sing}(x)\phi_{sing}(x')\right\rangle $
contains the entire singular piece of the TPF, both limits at the
R.H.S. are well defined. We denote the second limit by $\left\langle \phi^{2}\left(x\right)\right\rangle _{Analytic}$.
This quantity is well-defined and regular, and it is computed analytically
by summing/integrating over the WKB expressions for the large $\omega,l$
modes which comprise $\phi_{sing}$. The first term in the R.H.S.
is now expressed as a mode sum, and owing to its regularity the limit
may trivially be taken by replacing $x'$ by $x$. Putting it all
together one obtains 
\begin{multline}
\left\langle \phi^{2}\left(x\right)\right\rangle _{ren}=\hbar\int_{0}^{\infty}d\omega\sum_{l=0}^{\infty}\sum_{m=-l}^{l}\left|Y_{lm}\left(\theta,\varphi\right)\right|^{2}\left(\left|\bar{\psi}_{\omega l}\left(r\right)\right|^{2}-\left|\bar{\psi}_{\omega l}^{sing}\left(r\right)\right|^{2}\right)+\left\langle \phi^{2}\left(x\right)\right\rangle _{Analytic}.\label{eq: basic PS - phi Ren final}
\end{multline}
The mode sum should now converge (even though it is taken in coincidence),
owing to the subtraction of the singular piece. 

Using this method with splitting in the $t$ direction, $\left\langle \phi^{2}\right\rangle _{ren}$
(as well as $\left\langle T_{\alpha\beta}\right\rangle _{ren}$) was
calculated for the Schwarzschild case \cite{Candelas =000026 Howard - 1984 - phi2 Schwrazschild,Howard - 1984 - Tab Schwarzschild}
, and later also for a generic static spherically-symmetric spacetime
\cite{Anderson - 1990 - phi2 static spherically symmetric,Anderson - 1995 - Tab static spherically symmetric}. 

Besides the need to numerically compute the various mode functions
$\bar{\psi}_{\omega l}\left(r\right)$, this method also includes
a challenging analytical component: It requires a high-order WKB analysis.
As was mentioned in the Introduction, the presence of a turning point
makes this an extremely difficult task. To overcome these difficulties,
Candelas and Howard \cite{Candelas =000026 Howard - 1984 - phi2 Schwrazschild}
and others \cite{Anderson - 1990 - phi2 static spherically symmetric,Anderson - 1995 - Tab static spherically symmetric}
actually carried the analysis in the Euclidean sector.

\subsection{Remarks about the TPF and its mode sum\label{sub: Remarks-about-the-TPF}
\footnote{This subsection is somewhat remote from the main line of this paper,
and can be skipped at first reading.%
}}

\subsubsection{Caveats concerning the regularity of the TPF itself}

The point-splitting method is based on the presumption that the two-point
function is well-behaved as long as the two points are separated.
A few caveats are associated to this issue: First, even in flat spacetime,
the two-point function diverges when the separation is in a null direction.
Let us therefore assume, for the sake of simplicity, that the points
are separated in a timelike (or possibly spacelike) direction. The
second caveat is that in curved spacetime, assuming that the two points
are indeed separated by a timelike geodesic, if the (proper-time)
distance between $x$ and $x'$ is sufficiently large, there may also
be a null geodesic connecting these two points. For example, consider
the Schwarzschild spacetime and an approximately-static timelike geodesic
$\Gamma$ located in the asymptotic region very far from the BH. Let
$x$ be a point on $\Gamma$. There is a null geodesic which emanates
from $x$ and moves towards the BH (but with an appropriate miss),
makes a turn around the BH, then returns to $\Gamma$ and hits it
at a point $x_{1}$. It turns out that $\left\langle \phi\left(x\right)\phi\left(x'\right)\right\rangle $
develops a singularity at $x'=x_{1}$ \cite{Kay Radzikowski and Wald - 1996}.
In fact, along the timelike geodesic $\Gamma$ there is an infinite
discrete set of points $x_{n}$ which are connected to $x$ by null
geodesics that make $n$ turns around the BH before returning to $\Gamma$,
and $\left\langle \phi\left(x\right)\phi\left(x'\right)\right\rangle $
diverges at all points $x'=x_{n}$.

Note, however, that the problematic points $x_{n}$ are all located
far away from $x$, outside the normal neighborhood; and for the point-splitting
procedure only points $x'$ in the immediate neighborhood of $x$
are relevant. We shall thus restrict our attention now to points $x'$
in the close neighborhood of $x$, with a timelike (or alternatively
spacelike) separation between $x$ and $x'$. Then the TPF should
be well-behaved.

\subsubsection{Caveats concerning the convergence of the mode sum}

Naively one might expect that since the TPF is well behaved for a
short timelike or spacelike separation, the mode sum associated to
it should converge. It turns out, however, that the situation is more
subtle: The involved sum/integral usually fails to converge (in the
literal sense). Typically this failure to converge is associated with
undamped oscillations. 

This phenomenon, the failure of strict convergence, occurs already
in flat spacetime. The (non-) convergence situation may depend on
the direction of splitting, on the order of operations (summation
over $l,m$ and integration over $\omega$), and on the specific quantity
being calculated (whether it is the TPF, or a component of $\left\langle T_{\alpha\beta}\right\rangle $).
As an example, consider the calculation of $\left\langle \phi\left(x\right)\phi\left(x'\right)\right\rangle $
in Minkowski spacetime, using standard mode decomposition in spherical
harmonics and temporal modes $e^{-i\omega t}$. The mode functions
are given by 
\[
\bar{\psi}_{\omega l}\left(r\right)=\frac{1}{\sqrt{2r}}J_{l+1/2}\left(\omega r\right),
\]
where $J$ is the Bessel function of the first kind. Consider now
a point $x$ located at some $r>0$, and a point $x'$ displaced in
the $t$ direction by an amount $t'-t=\Delta t.$ The sum over $l,m$
(for a given $\omega$) then converges, and one finds that 
\[
\sum_{l=0}^{\infty}\sum_{m=-l}^{l}\left|Y_{lm}\left(\theta,\varphi\right)\right|^{2}\left|\bar{\psi}_{\omega l}\left(r\right)\right|^{2}=\frac{1}{4\pi^{2}}\omega.
\]
The mode-sum expression for the TPF then becomes 
\begin{equation}
\left\langle \phi\left(x\right)\phi\left(x'\right)\right\rangle =\frac{\hbar}{4\pi^{2}}\int_{0}^{\infty}\omega e^{i\omega\Delta t}d\omega.\label{eq:  A3}
\end{equation}
The problem is that this integral does not converge in the usual sense
(even conditionally), due to the growing oscillations at large $\omega$. 

In this example of pure $t$-splitting, in which we first sum over
$m,l$, it was the integral over $\omega$ that failed to converge.
In another application of point splitting, in which the splitting
is in both $t$ and $\theta$, and in which one first integrates over
$\omega$ and sums over $m,l$ afterward, one finds that this time
the sum over $l$ fails to converge, again due to growing oscillations.
\footnote{In addition, the integral over $\omega$ converges for the TPF but
only conditionally, and fails to converge even conditionally for $T_{tt}$,
due to growing oscillations.%
}

This situation, of non-converging oscillatory integrals (or sums over
$l$), is fairly common in various QFT calculations. The common practice
(which may be justified by several arguments) is to contend that the
large-$\omega$ oscillations should be damped in some appropriate
manner.

\subsubsection{Generalized integral }

This situation motivates us to introduce the notion of generalized
integral, which properly incorporates oscillation damping. We should
emphasize that all the $\omega$ integrals in this paper are in fact
generalized integrals.

We shall consider here two specific procedures of oscillation damping,
which yield two (mutually consistent) definitions of generalized integral.

\paragraph{Abel-summation integral:}

This is a commonly-used method for giving a meaning for such oscillatory
integrals. The Abel-summed integral is defined as
\begin{equation}
\int_{0}^{\infty(A)}h(\omega)d\omega\equiv\lim_{\epsilon\rightarrow0_{+}}\int_{0}^{\infty}e^{-\epsilon\omega}h(\omega)d\omega\label{eq:  A72}
\end{equation}
 --- provided of course that the integral at the R.H.S. is well defined
for $\epsilon>0$, and the limit $\epsilon\rightarrow0_{+}$ exists.
For example, with Abel summation, the integral in Eq. (\ref{eq:  A3})
reads $-1/\Delta t^{2}$ --- yielding the standard expression $-\hbar/\left(4\pi^{2}\Delta t^{2}\right)$
for the flat-space TPF {[}cf. Eq. (\ref{eq: basic PS - The counter-terms GDS}){]}. 

A crucial property of the Abel-summation integral is that it is consistent
with the standard integral. Namely, whenever the function $h(\omega)$
is integrable in the strict sense, its Abel-summation integral coincides
with the standard integral of this function.

\paragraph*{Self-cancellation integral:}

For the calculation of $\left\langle \phi^{2}\right\rangle _{ren}$
in a black-hole spacetime we shall have to carry (generalized) integral
of oscillatory functions $h(\omega)$ that we determine numerically.
In such a case, this procedure of Abel summation --- which must now
be implemented numerically --- is fairly inconvenient. Furthermore,
in practice we only determine $h(\omega)$ in a restricted range $0<\omega<\omega_{max}$,
which makes the Abel-summation integral even harder to implement.
\footnote{To this end one would have to generalize the definition of the Abel-summation
integral, so as to combine the two (non-commuting) limits $\epsilon\rightarrow0$
and $\omega_{max}\rightarrow\infty$ in an appropriate manner. And
it turns out that the convergence of this generalized Abel integral
with increasing $\omega_{max}$ is rather slow. %
} We therefore find it much more convenient to use another concept
of generalized integral, which we name \emph{self-cancellation} (of
the oscillations). This type of generalized integral is applicable
whenever the oscillations have well-defined frequencies---which is
indeed the situation in our problem (see Appendix \ref{sec: Appendix-B}).
To formulate this concept, we first define the integral function 
\begin{equation}
H(\omega)\equiv\int_{0}^{\omega}h(x)dx.\label{eq:  H1}
\end{equation}
The standard integral may then be expressed as 
\[
\int_{0}^{\infty}h(\omega)d\omega=\lim_{\omega\rightarrow\infty}H(\omega)
\]
(whenever this limit exists). Instead, our self-cancellation generalized
integral is defined as
\begin{equation}
\int_{0}^{\infty(sc)}h(\omega)d\omega\equiv\lim_{\omega\rightarrow\infty}\left[\frac{H(\omega)+H(\omega+\lambda/2)}{2}\right].\label{eq:  A22}
\end{equation}
Here, $\lambda$ denotes the ``wave-length'' of the oscillation
in $h(\omega)$ (which is also inherited by $H$). For example, in
Eq. (\ref{eq:  A3}) the oscillatory factor is $e^{i\omega\Delta t}$,
hence the period of oscillation is $\lambda=2\pi/\Delta t$. The idea
is simple: The non-oscillating piece of $H(\omega)$ is unaffected
by this averaging, but the oscillatory piece will be very effectively
annihilated by such averaging with half-wavelength shift.

As a simplest example, consider the case $h(\omega)=\sin(\omega L)$.
Then $H(\omega)=\left[1-\cos(\omega L)\right]/L$, and obviously $\lambda=2\pi/L$.
Clearly $H(\omega)$ fails to have a limit $\omega\rightarrow\infty$.
Yet the self-cancellation integral is perfectly well-defined: The
term in squared brackets in Eq. (\ref{eq:  A22}) is simply $1/L$,
entirely independent of $\omega$. This example demonstrates the potential
of the self-cancellation method to yield extremely fast convergence
in $\omega$. This last property is important, especially because
in an actual calculation we have to determine $h(\omega)$ numerically,
and we do so in a restricted range of $\omega$. 

For later convenience we re-formulate this notion of self-cancellation
integral as follows: 
\begin{equation}
\int_{0}^{\infty(sc)}h(\omega)d\omega\equiv\lim_{\omega\rightarrow\infty}T_{\lambda}\left[H(\omega)\right],\label{eq:  A27}
\end{equation}
where $T_{\lambda}$ is the ``self-cancellation operation'' defined
by 
\[
T_{\lambda}\left[f(\omega)\right]\equiv\frac{f(\omega)+f(\omega+\lambda/2)}{2}.
\]

In the actual calculation of $\left\langle \phi^{2}\right\rangle $
we shall have to repeat the self-cancellation operation several times
(for several different oscillation frequencies). In Appendix \ref{sec: Appendix-A}
we shall introduce the ``multiple self-cancellation operation''
$T_{*}$, formed by combining several $T_{\lambda}$ operations. 

It is easy to show that the notion of self-cancellation integral is
fully consistent with the standard integral --- whenever the latter
is well defined. Furthermore, the self-cancellation integral is also
fully consistent with the Abel-summation integral, in the following
sense: If the self-cancellation integral converges, then the Abel
integral converges too, and the two generalized integrals yield the
same result. Note, however, that the Abel summation method is more
general than self-cancellation. Namely, there are functions $h(\omega)$
for which the Abel-summed integral is well defined but the self-cancellation
integral is not (that is, the limit in Eq. (\ref{eq:  A22}) is non-existent).
Nevertheless, for the functions $h(\omega)$ involved in the analysis
below, the self-cancellation integral is well-defined and extremely
powerful.

\section{Our new method: The $t$-splitting variant \label{sec:The-new-scheme}}

As was already noted in the Introduction, since our ultimate goal
is to address dynamical background metrics as well, we shall carry
the analysis directly in the Lorentzian sector. Hence, due to the
inevitable presence of a turning point (and also due to the PDE nature
of the time-dependent mode equation), we shall avoid using WKB expansion
in our method. Instead we extract the required information about the
high-frequency modes of the field directly from the counter-term (\ref{eq: basic PS - The counter-terms GDS}).

We shall present here the $t$-splitting variant, which requires a
time-translation Killing field. Although this variant should be applicable
to a rather generic stationary asymptotically-flat background, we
shall restrict our attention first to the more specific case of static
spherically-symmetric background, for the sake of simplicity. Then
in Sec. \ref{sub: Our method - Non spherical symmetric} we outline
the generalization of the $t$-splitting formulation beyond spherical
symmetry (and beyond staticity), to a more generic stationary asymptotically-flat
background.

\subsection{Spherically-symmetric static background\label{sub: Our method -  spherical symmetric}}

We split the points in the $t$ direction, namely 
\begin{equation}
x=(t,r,\theta,\varphi),\, x'=(t+\varepsilon,r,\theta,\varphi).\label{eq: Our scheme - the points x,x' in the t split}
\end{equation}
The TPF then takes the form 
\begin{equation}
\left\langle \phi\left(x\right)\phi\left(x'\right)\right\rangle =\hbar\int_{0}^{\infty}d\omega\, e^{i\omega\varepsilon}\sum_{l=0}^{\infty}\sum_{m=-l}^{l}\left|Y_{lm}\left(\theta,\varphi\right)\right|^{2}\left|\bar{\psi}_{\omega l}\left(r\right)\right|^{2}.\label{eq: A1}
\end{equation}
As was already mentioned above, the sum over $l,m$ converges, and
we denote it by $F\left(\omega,r\right)$. In fact, the sum of $|Y_{lm}|^{2}$
over $m$ yields $(2l+1)/4\pi$, and therefore 
\begin{equation}
F\left(\omega,r\right)=\sum_{l=0}^{\infty}\frac{2l+1}{4\pi}\left|\bar{\psi}_{\omega l}\left(r\right)\right|^{2}.\label{eq: Our scheme - F defintion}
\end{equation}
This function is to be computed numerically. The TPF now reduces to
\begin{equation}
\left\langle \phi\left(x\right)\phi\left(x'\right)\right\rangle =\hbar\int_{0}^{\infty}F\left(\omega,r\right)e^{i\omega\varepsilon}d\omega.\label{eq: Our scheme - phi2 splitted}
\end{equation}
(Note that in the coincidence limit $\varepsilon\rightarrow0$ this
integral would diverge. However, the oscillatory factor $e^{i\omega\varepsilon}$
regularizes it.) Equation (\ref{eq: basic PS - phi Ren limit}) now
reads 
\begin{equation}
\left\langle \phi^{2}\left(x\right)\right\rangle _{ren}=\lim_{\varepsilon\to0}\left[\hbar\int_{0}^{\infty}F\left(\omega,r\right)e^{i\omega\varepsilon}d\omega-G_{DS}\left(\varepsilon\right)\right],\label{eq:  A2}
\end{equation}
where by $G_{DS}\left(\varepsilon\right)$ we refer to $G_{DS}\left(x,x'\right)$
with $x'$ given by Eq. (\ref{eq: Our scheme - the points x,x' in the t split}).
Note that for a given $x$, $\sigma$ is uniquely determined by $\varepsilon$.
By a fairly straightforward Taylor expansion of the geodesic equation
(and $\sigma$) in $\varepsilon$, one finds that $G_{DS}$ in Eq.
(\ref{eq: basic PS - The counter-terms GDS}) takes the general form
 
\begin{equation}
\frac{1}{\hbar}G_{DS}\left(x,x'\right)=a\left(r\right)\varepsilon^{-2}+c\left(r\right)\left[\ln\left(\varepsilon\mu\right)+\gamma-\frac{i\pi}{2}\right]+d\left(r\right)+O(\varepsilon),\label{eq: Our scheme - The counter-terms}
\end{equation}
where $a(r),c(r),d(r)$ are certain (real) functions that depend on
the background metric and the parameters of the field. %
\footnote{The term $-i\pi/2$ appears in the brackets because $\sigma$ is negative.%
} The explicit form of these functions is not important in the present
discussion (although it is certainly needed for the actual calculation
of $\left\langle \phi^{2}\right\rangle _{ren}$). 

To proceed, we now decompose the $\varepsilon$-dependent terms in
$G_{DS}$ using the Laplace transform. We have the following identities:
\begin{equation}
\varepsilon^{-2}=-\int_{0}^{\infty}\omega e^{i\omega\varepsilon}d\omega,\label{eq: Our scheme - transformation of the counter terms - a}
\end{equation}

\begin{equation}
\ln\left(\varepsilon\mu\right)=-\int_{0}^{\infty}\frac{1}{\omega+\mu}e^{i\omega\varepsilon}d\omega+\left(\frac{i\pi}{2}-\gamma\right)\,+\, O\left(\varepsilon\ln\varepsilon\right).\label{eq: Our scheme - transformation of the counter terms - c}
\end{equation}
Inserting these identities into Eqs. (\ref{eq:  A2},\ref{eq: Our scheme - The counter-terms})
we obtain 
\begin{equation}
\left\langle \phi^{2}\left(x\right)\right\rangle _{ren}=\hbar\lim_{\varepsilon\to0}\int_{0}^{\infty}F_{reg}\left(\omega,r\right)e^{i\omega\varepsilon}d\omega-\hbar\, d\left(r\right)\label{eq: Our scheme - phi analytic}
\end{equation}
where 
\begin{equation}
F_{reg}\left(\omega,r\right)\equiv F\left(\omega,r\right)-F_{sing}\left(\omega,r\right),\label{eq:  A5}
\end{equation}
 and 
\begin{equation}
F_{sing}\left(\omega,r\right)\equiv-a\left(r\right)\omega-c\left(r\right)\frac{1}{\omega+\mu}.\label{eq:  A6}
\end{equation}

Consider now the R.H.S. of Eq. (\ref{eq: Our scheme - phi analytic}).
If the integral of $F_{reg}\left(\omega,r\right)$ converges (in either
the strict sense or the generalized sense), we can interchange the
limit and integration, and get rid of the $\varepsilon\to0$ limit
altogether. We should expect the convergence of the integral of $F_{reg}\left(\omega,r\right)$,
because the singular piece $F_{sing}\left(\omega,r\right)$ has already
been removed from $F\left(\omega,r\right)$. In fact, we find that
in the Schwarzschild case the integral of $F_{reg}\left(\omega,r\right)$
indeed converges, although only in the generalized sense due to oscillations
(see next section). We were unable to \emph{prove} the (even generalized)
convergence of the integral of $F_{reg}\left(\omega,r\right)$, but
nevertheless since this convergence is naturally expected, and since
the Schwarzschild example confirms this expectation, we shall hereafter
\emph{assume} that this integral indeed converges in the generalized
sense. %
\footnote{Note that there is no much risk in making such an assumption, because
if for a certain background metric this assumption turns out to be
false, then the attempt to integrate $F_{reg}$ will demonstrate this
non-convergence right away. %
}

We therefore write our final result as  
\begin{equation}
\left\langle \phi^{2}\left(x\right)\right\rangle _{ren}=\hbar\int_{0}^{\infty}F_{reg}\left(\omega,r\right)d\omega-\hbar\, d\left(r\right)\label{eq: Our scheme - phi analytic-Final}
\end{equation}
where, recall, the integral over $\omega$ is a generalized one (as
defined in Sec. 2.2). The implementation of this generalized integral
is demonstrated in Sec. \ref{sec: Calc in Schwarzschild } for the
Schwarzschild case.

\subsection{The general stationary case\label{sub: Our method - Non spherical symmetric}}

As was already pointed out above, our $t$-splitting method does not
require the background metric to be spherically symmetric or even
static: It should be applicable to a generic stationary asymptotically-flat
background. Here we outline this extension. 

On account of asymptotic flatness, we now choose our coordinates ($t,r,\theta,\varphi)$
such that the large$-r$ asymptotic metric still takes its standard
Minkowski form $-dt^{2}+dr^{2}+r^{2}d\Omega^{2}$. %
\footnote{Furthermore we choose our coordinates such that $g_{tt}$ takes the
standard weak-field form $-1+2M/r+O(1/r)^{2}$, where here $M$ denotes
the system's asymptotic mass. %
} Due to lack of spherical symmetry, the expression (\ref{eq: basic PS - expression for f})
for the field modes is now replace by
\[
f_{\omega lm}\left(x\right)=e^{-i\omega t}\tilde{\psi}_{\omega lm}\left(r,\theta,\varphi\right),
\]
where $\tilde{\psi}_{\omega lm}\left(r,\theta,\varphi\right)$ is
a set of solutions to the ($\omega$-dependent) spatial part of the
field equation, which is now a PDE (in $r,\theta,\varphi$) rather
than ODE. These solutions are required to be regular everywhere, and
to satisfy the large-$r$ boundary condition 
\[
\tilde{\psi}_{\omega lm}\left(r,\theta,\varphi\right)=(r\sqrt{4\pi\omega})^{-1}e^{-i\omega r_{*}}Y_{lm}\left(\theta,\varphi\right)+\tilde{\psi}_{\omega lm}^{out}\left(r,\theta,\varphi\right),\qquad\qquad(r\rightarrow\infty)
\]
where $\tilde{\psi}_{\omega lm}^{out}$ denotes the reflected field
which is $e^{i\omega r_{*}}/r$ times some function of $\theta$ and
$\varphi$. %
\footnote{Here $r_{*}$ is to be regarded as the standard function of $r$ given
in Eq. (\ref{eq:  r-star}).%
} Once the modes $f_{\omega lm}\left(x\right)$ were defined, the expression
(\ref{eq: basic PS - field decomposition to a,ad}) for the field
operator is unchanged. 

The TPF now takes the form 
\begin{equation}
\left\langle \phi\left(x\right)\phi\left(x'\right)\right\rangle =\hbar\int_{0}^{\infty}d\omega\, e^{i\omega\varepsilon}\sum_{l=0}^{\infty}\sum_{m=-l}^{l}\left|\tilde{\psi}_{\omega lm}\left(r,\theta,\varphi\right)\right|^{2}.\label{eq: General stationary - TPF}
\end{equation}
It is important to recall that even though the metric is not spherically
symmetric, the sums over $m,l$ should still converge (for a given
$\omega$): Due to asymptotic flatness, at large $r$ there will be
a centrifugal barrier $\approx l(l+1)/r^{2}$ in the effective potential,
just like in Minkowski, preventing the penetration of modes with too
large $l$. Thus we can again define 
\begin{equation}
F\left(\omega,r,\theta,\varphi\right)=\sum_{l=0}^{\infty}\sum_{m=-l}^{l}\left|\tilde{\psi}_{\omega lm}\left(r,\theta,\varphi\right)\right|^{2},\label{eq: General stationary - F definition}
\end{equation}
and the TPF still takes the form (\ref{eq: Our scheme - phi2 splitted})
{[}although now with $F\left(\omega,r,\theta,\varphi\right)${]}.
The calculation now proceeds just as in the spherically symmetric
case --- except that all the quantities in Sec. (\ref{sub: Our method -  spherical symmetric})
that were dependent on $r$ only, now depend on $\theta$ and $\varphi$
as well. The final result is  
\begin{equation}
\left\langle \phi^{2}\left(x\right)\right\rangle _{ren}=\hbar\int_{0}^{\infty}F_{reg}\left(\omega,r,\theta,\varphi\right)d\omega-\hbar\, d\left(r,\theta,\varphi\right)\label{eq: Our scheme - phi analytic-Final-gen}
\end{equation}
with
\begin{equation}
F_{reg}\left(\omega,r,\theta,\varphi\right)\equiv F\left(\omega,r,\theta,\varphi\right)-F_{sing}\left(\omega,r,\theta,\varphi\right),\label{eq:  A5-gen}
\end{equation}
 and 
\begin{equation}
F_{sing}\left(\omega,r,\theta,\varphi\right)\equiv-a\left(r,\theta,\varphi\right)\omega-c\left(r,\theta,\varphi\right)\frac{1}{\omega+\mu}.\label{eq:  A6-gen}
\end{equation}

It should be pointed out that although the choice of the coordinates
$r,\theta,\varphi$ is certainly non-unique (due to lack of spherical
symmetry), the resultant mode decomposition is still unique. This
is because of asymptotic flatness (and the standard weak-field metric
that we require our $t,r,\theta,\varphi$ coordinates to satisfy at
large $r$), and because the mode functions $\tilde{\psi}_{\omega lm}\left(r,\theta,\varphi\right)$
are defined through their asymptotic form at $r\rightarrow\infty$.
The unique mode decomposition in turn leads to a unique Fock space
associated to it, and to a well-defined vacuum state. 

Note, however, that in the above formulation we implicitly assumed
a stationary background metric with no past horizon (e.g. a spinning
star), hence it was sufficient to construct the ``in'' modes. In
the case of eternal non-spherical BH the situation becomes more subtle,
because now we also need to specify the ``up'' modes. Here asymptotic
flatness will not be of much help, because these modes are to be defined
by boundary conditions at the past horizon. With the lack of spherical
symmetry and staticity, one still needs to figure out how to make
a unique mode decomposition and to obtain a ``natural'' vacuum state.
\footnote{Furthermore, with an arbitrary choice of $r,\theta,\varphi$ coordinates,
and with a corresponding arbitrary construction of the set of ``up''
modes,  we have no guarantee that the resultant ``vacuum state''
would at all be a well-defined Hadamard state. In this regard, we
should mention the observation that there is no Hadamard state that
respects the symmetries of Kerr spacetime and regular everywhere.
\cite{Kay and Wald - 1991,Ottewill - 2000} %
} 

The implementation of this method in the non-spherical case is of
course technically more challenging (even for a non-BH background),
because now the mode functions $\tilde{\psi}_{\omega lm}$ which comprise
$F\left(\omega,r,\theta,\varphi\right)$ are to be obtained by numerically
solving PDEs rather than just ODEs.

\section{Calculation of $\left\langle \phi^{2}\right\rangle _{ren}$ in Schwarzschild
\label{sec: Calc in Schwarzschild }}

Using the method presented in the last section, we compute $\left\langle \phi^{2}\right\rangle _{ren}$
in the exterior region of Schwarzschild spacetime, in the Boulware
vacuum state, for a minimally coupled massless scalar field. The Schwarzschild
metric is
\begin{equation}
ds^{2}=-\left(1-\frac{2M}{r}\right)dt^{2}+\left(1-\frac{2M}{r}\right)^{-1}dr^{2}+r^{2}d\Omega^{2}.\label{eq: Schwarzschild calc - Metric}
\end{equation}
In this metric the quantum field $\phi\left(x\right)$ can be expanded
in the form presented in Eqs. (\ref{eq: basic PS - field decomposition to a,ad})-(\ref{eq: basic PS - expression for f}),
where $\bar{\psi}_{\omega l}\left(r\right)$ is conveniently recast
as 
\begin{equation}
\bar{\psi}_{\omega l}\left(r\right)=\frac{1}{r\sqrt{4\pi\omega}}\psi_{\omega l}\left(r\right),\label{eq: Schwarzschild calc - psi bar and psi relation}
\end{equation}
and $\psi_{\omega l}\left(r\right)$ satisfies the radial equation
\begin{equation}
\frac{d^{2}\psi_{\omega l}\left(r\right)}{dr_{*}^{2}}=-\left[\omega^{2}-V_{l}\left(r\right)\right]\psi_{\omega l}\left(r\right).\label{eq: Schwarzschild calc - radial equation}
\end{equation}
Henceforth, $r_{*}$ will denote the tortoise coordinate given by
\begin{equation}
r_{*}=r+2M\ln\left(\frac{r}{2M}-1\right),\label{eq:  r-star}
\end{equation}
and 
\[
V_{l}\left(r\right)=\left(1-\frac{2M}{r}\right)\left[\frac{l\left(l+1\right)}{r^{2}}+\frac{2M}{r^{3}}\right]
\]
is the effective potential.

The general solution of Eq. (\ref{eq: Schwarzschild calc - radial equation})
(for given $\omega,l$) is spanned by two basic solutions, that we
denote $\psi_{\omega l}^{in}\left(r\right)$ and $\psi_{\omega l}^{up}\left(r\right)$.
The boundary conditions for these two basic solutions are taken to
be
\begin{gather}
\psi_{\omega l}^{in}\left(r\right)=\left\{ \begin{array}[t]{cc}
\tau_{\omega l}\, e^{-i\omega r_{*}}, & r_{*}\to-\infty\\
e^{-i\omega r_{*}}+\rho_{\omega l}\, e^{i\omega r_{*}},\,\,\,\,\,\, & r_{*}\to\infty
\end{array}\right.\nonumber \\
\psi_{\omega l}^{up}\left(r\right)=\left\{ \begin{array}[t]{cc}
e^{i\omega r_{*}}+\tilde{\rho}_{\omega l}\, e^{-i\omega r_{*}},\,\,\,\,\,\, & r_{*}\to-\infty\\
\tau_{\omega l}\, e^{i\omega r_{*}}, & r_{*}\to\infty
\end{array}\right.\label{eq: Schwarzschild calc - Basic solutions bounadry conditions}
\end{gather}
where $\tau_{\omega l}$, $\rho_{\omega l}$, and $\tilde{\rho}_{\omega l}=-\rho_{\omega l}^{*}\tau_{\omega l}/\tau_{\omega l}^{*}$
represent the transmission and reflection amplitudes. These two basic
solutions are properly normalized and mutually orthogonal. 

The presence of two independent modes for each $\omega lm$ (as opposed
to a single such mode in e.g. Minkowski) requires a slight modification
of the formalism above: We now have \emph{two} sets of annihilation
operators, $a_{\omega lm}^{in}$ and $a_{\omega lm}^{up}$---as well
as their conjugate operators $a_{\omega lm}^{\dagger in}$, $a_{\omega lm}^{\dagger up}$.
Correspondingly, in Eq. (\ref{eq: basic PS - field decomposition to a,ad}),
in addition to the integral over $\omega$ and the sum over $l$ and
$m$, we also have to sum over the separate contributions of the ``in''
and ``up'' modes. We shall consider here the B\emph{oulware vacuum},
namely the quantum state annihilated by all the operators $a_{\omega lm}^{up}$
as well as $a_{\omega lm}^{in}$. Revisiting the analysis of the previous
sections, one finds that everything remains intact, except that all
the equations that involve summation over $l,m$ should now also include
a summation over the ``in'' and ``up'' contributions. Correspondingly,
Eq. (\ref{eq: Our scheme - F defintion}) is now replaced by
\begin{equation}
F\left(\omega,r\right)=\sum_{l=0}^{\infty}\frac{2l+1}{4\pi}\left(\left|\bar{\psi}_{\omega l}^{in}\left(r\right)\right|^{2}+\left|\bar{\psi}_{\omega l}^{up}\left(r\right)\right|^{2}\right),\label{eq: Schwarzschild calc - F omega}
\end{equation}
but otherwise the results of the previous section, and in particular
Eqs. (\ref{eq:  A5},\ref{eq:  A6},\ref{eq: Our scheme - phi analytic-Final}),
are unaffected. 

Since the Schwarzschild metric is a vacuum solution, and since we
are dealing with a massless field, the counter-term (\ref{eq: basic PS - The counter-terms GDS})
now reduces to 
\begin{equation}
G_{DS}\left(x,x'\right)=\frac{\hbar}{8\pi^{2}\sigma}.\label{eq: Counter-term_Schwarzschild}
\end{equation}
$\sigma$ is related to $\varepsilon\equiv t'-t$ through the proper-time
$\tau$ of the short geodesic connecting $x$ to $x'$, via $\sigma=-\tau^{2}/2$.
By conducting a second-order expansion of the geodesic equation we
obtain 
\begin{equation}
\sigma\left(\varepsilon\right)=-\frac{1-2M/r}{2}\varepsilon^{2}-\frac{M^{2}\left(1-2M/r\right)}{24r^{4}}\varepsilon^{4}+O(\varepsilon^{5}),\label{eq: Schwarzschild calc - sigma}
\end{equation}
and correspondingly 
\begin{equation}
\frac{1}{\hbar}G_{DS}\left(x,x'\right)=-\frac{1}{4\pi^{2}\left(1-2M/r\right)}\varepsilon^{-2}+\frac{M^{2}}{48\pi^{2}r^{4}\left(1-2M/r\right)}+O(\varepsilon).\label{eq: Schwarzschild calc - counter terms}
\end{equation}
Comparing this to Eq. (\ref{eq: Our scheme - The counter-terms})
we find that
\begin{equation}
a(r)=-\frac{1}{4\pi^{2}\left(1-2M/r\right)}\:,\;\;\; d(r)=\frac{M^{2}}{48\pi^{2}r^{4}\left(1-2M/r\right)}\:,\;\;\; c(r)=0.\label{eq:  A10}
\end{equation}
Therefore in the Schwarzschild case Eqs. (\ref{eq:  A5},\ref{eq:  A6},\ref{eq: Our scheme - phi analytic-Final})
reduce to

\begin{equation}
\left\langle \phi^{2}\left(x\right)\right\rangle _{ren}=\hbar\int_{0}^{\infty}F_{reg}\left(\omega,r\right)d\omega-\hbar\, d\left(r\right).\label{eq:  A13}
\end{equation}
with
\begin{equation}
F_{reg}\left(\omega,r\right)=F\left(\omega,r\right)+a\left(r\right)\omega.\label{eq:  A12}
\end{equation}

Summarizing the analytical part of the calculation, $\left\langle \phi^{2}\left(x\right)\right\rangle _{ren}$
is given by Eq. (\ref{eq:  A13}) along with Eqs. (\ref{eq:  A12}),
(\ref{eq:  A10}) and (\ref{eq: Schwarzschild calc - F omega}). Note
that basically this expression for Schwarzschild was already obtained
by Candelas, \cite{Candelas - 1979 - phi2 Schwarzschild}, but here
we also complete the calculation by implementing the numerical part
as well (and by doing so, we encounter the oscillations problem and
address it).

\subsection{Numerical implementation}

We have numerically solved the radial equation for $\psi_{\omega l}^{in}\left(r\right)$
and $\psi_{\omega l}^{up}\left(r\right)$, using the standard MATHEMATICA
numerical ODE solver, in the domain $0<\omega<3$, at a set of $\omega$
values with a uniform separation $d\omega=1/300$. Hereafter, we use
units in which the BH mass is $M=1$ (in addition to $C=G=1$), hence
$\omega$ as well as $r$ are dimensionless. 

For a given $\omega$, the contribution of the different $l$ modes
to $F\left(\omega,r\right)$ starts to decay exponentially fast beyond
a certain $l$ value, typically of order $\sim\omega r\left(1-2M/r\right)^{-1/2}$.
(This decay may be interpreted as tunneling into the potential barrier.)
Correspondingly, for each $\omega$ value, we truncate the sum (\ref{eq: Schwarzschild calc - F omega})
at an $l$ value where the contribution becomes negligible ($<10^{-10})$.
Then from $F\left(\omega,r\right)$ we construct the regularized function
$F_{reg}\left(\omega,r\right)$ according to Eq. (\ref{eq:  A12}). 

\begin{figure}
\subfloat[The numerically calculated $F\left(\omega,r=6\right)$ in the Schwarzschild
case.\label{fig:1a}]{\begin{centering}
\includegraphics[bb=90bp 260bp 500bp 580bp,clip,scale=0.48]{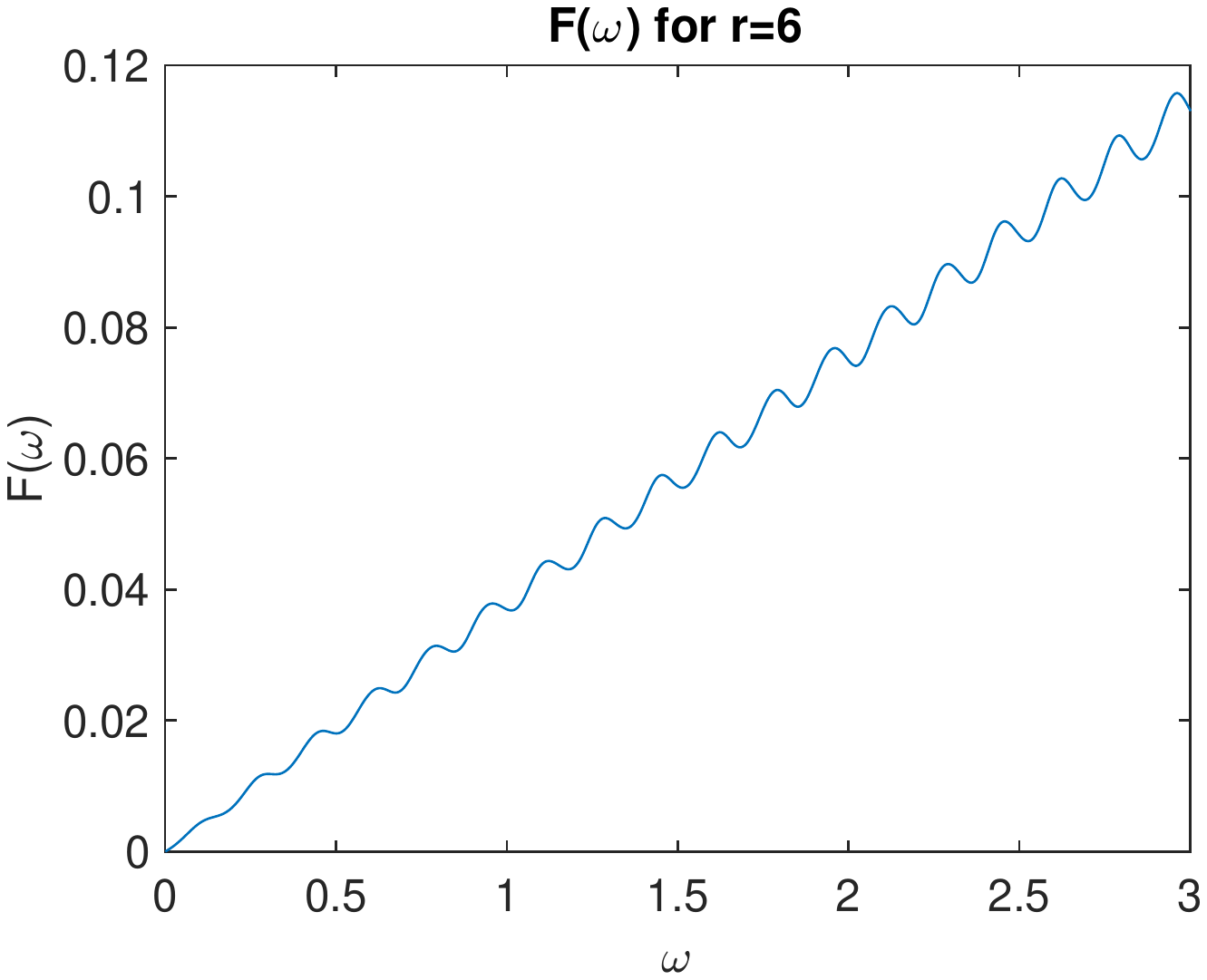}
\par\end{centering}

}\enskip{}\hfill{}\enskip{}\subfloat[$F_{reg}\left(\omega,r=6\right)$, which is the result of subtracting
the linearly-diverging piece $a\left(r\right)\omega$ from $F\left(\omega,r\right)$.\label{fig:1b}]{\begin{centering}
\includegraphics[bb=90bp 260bp 500bp 580bp,clip,scale=0.48]{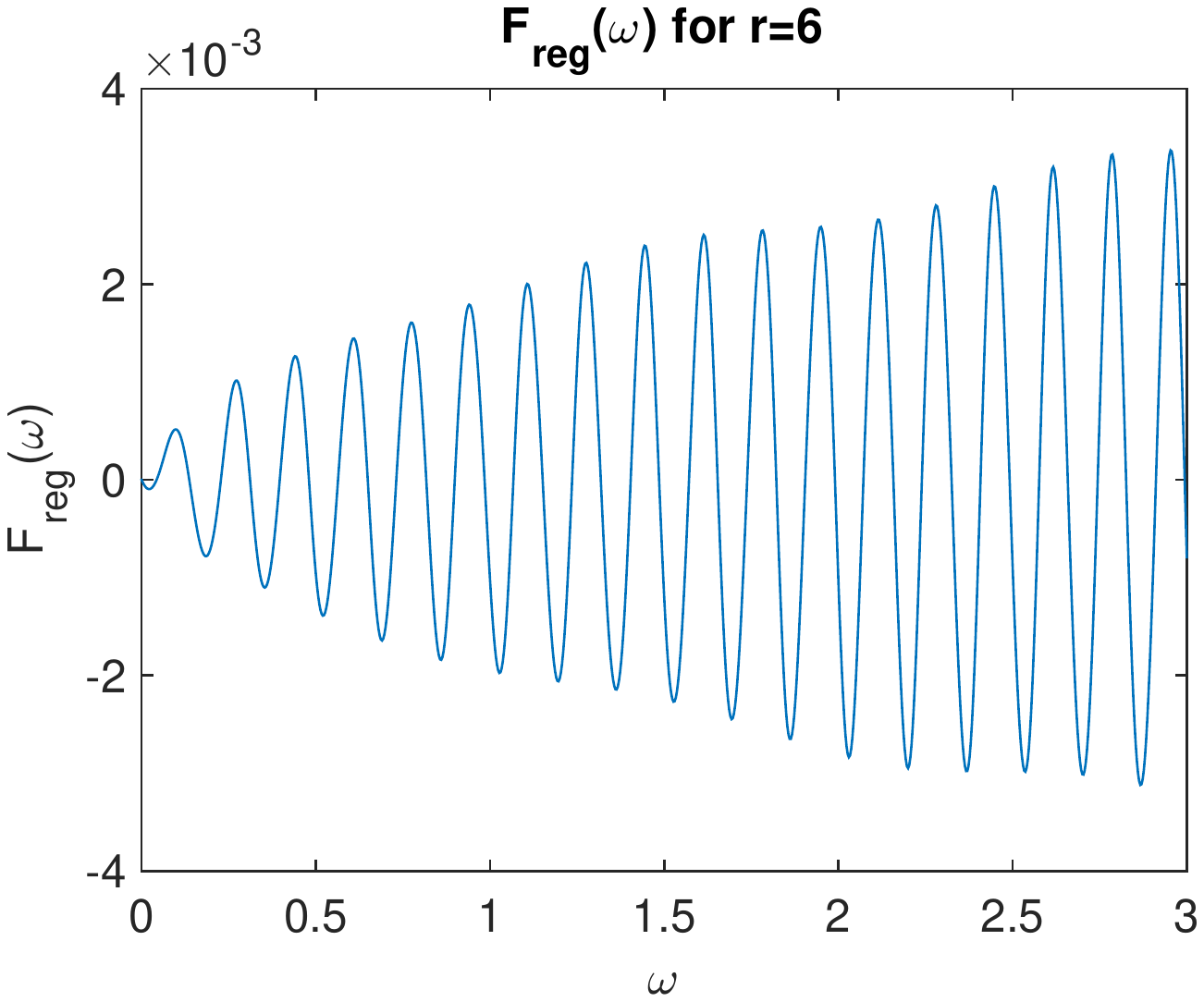}
\par\end{centering}

}

\protect\caption{}
\end{figure}

Figure \ref{fig:1a} displays $F\left(\omega,r\right)$ for $r=6$,
which we choose here as our representative $r$ value. Then Fig. \ref{fig:1b}
displays the regularized function $F_{reg}\left(\omega,r\right)$,
obtained from $F\left(\omega,r\right)$ by removing the linear piece
$a\left(r\right)\omega$. Clearly, the linear divergence has been
removed, but there remain oscillations that grow as $\omega^{1/2}$,
which we address below. The origin of these oscillations (the aforementioned
connecting null geodesics), and the determination of their frequencies,
are discussed in Appendix \ref{sec: Appendix-B}. 

\begin{figure}
\begin{centering}
\includegraphics[bb=90bp 260bp 500bp 580bp,clip,scale=0.6]{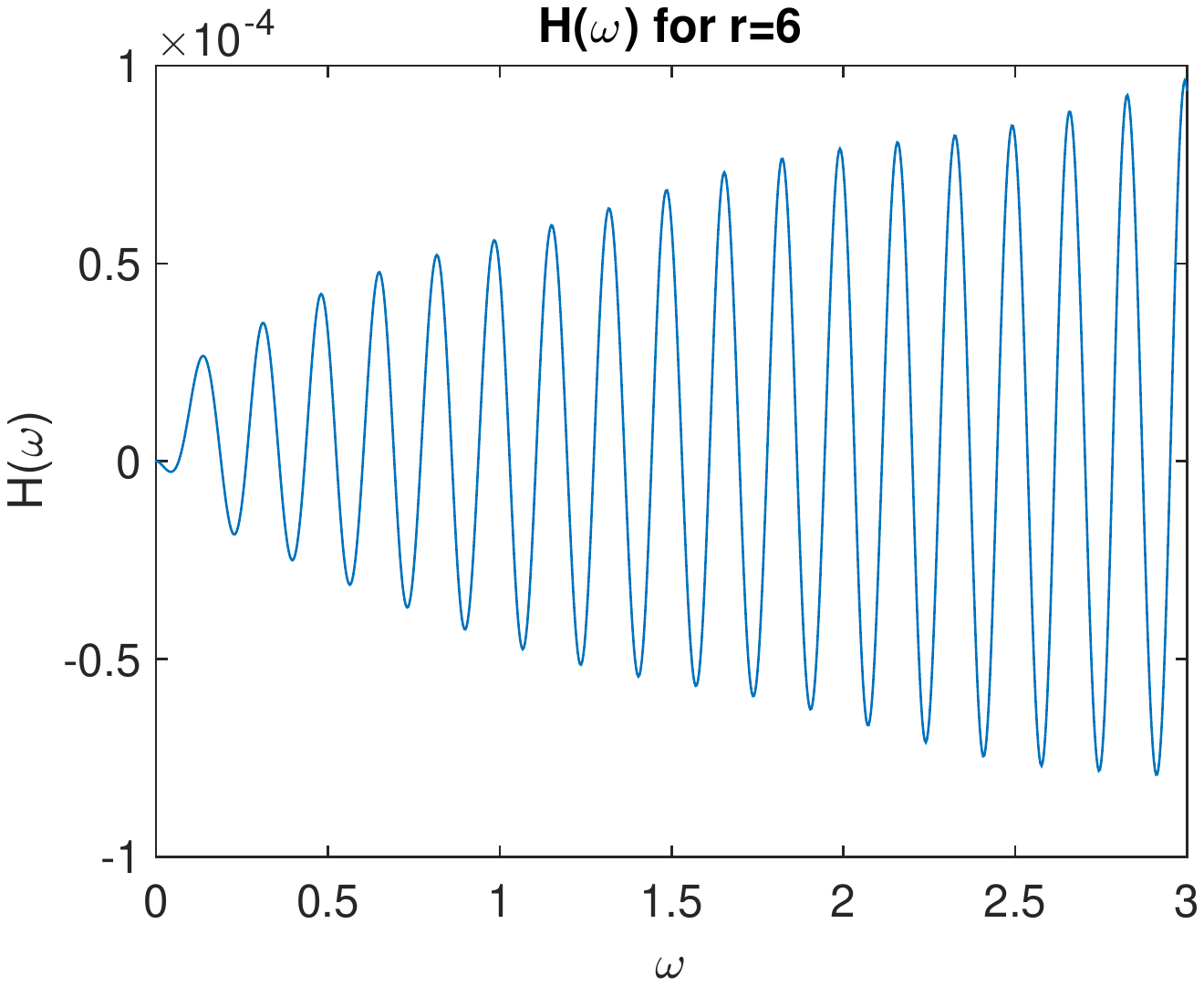}
\par\end{centering}

\protect\caption{The integral function $H(\omega,r=6)$ (that is, the integral of $F_{reg}$
from zero to $\omega$). \label{fig:2}}

\end{figure}

To calculate $\left\langle \phi^{2}\left(x\right)\right\rangle _{ren}$
we need the generalized integral of $F_{reg}\left(\omega,r\right)$
from $\omega=0$ to infinity, see Eq. (\ref{eq:  A13}). To this end
we define the integral function 
\begin{equation}
H(\omega)\equiv\int_{0}^{\omega}F_{reg}\left(\omega',r\right)d\omega'.\label{eq:  A39}
\end{equation}
The strict integral of $F_{reg}$ would of course correspond to the
limit $\omega\rightarrow\infty$ of $H(\omega)$. Figure \ref{fig:2}
displays $H(\omega)$, and makes it clear that this limit does not
exist, due to the growing oscillations (which were inherited directly
from $F_{reg}$). We therefore have to resort to the generalized integral
instead, as discussed in Sec. \ref{sub: Remarks-about-the-TPF}. The
Abel-summation integral (\ref{eq:  A72}) is well defined in this
case. However, since we know the precise frequency of oscillations,
it is much more convenient and more efficient to employ the self-cancellation
integral (which is fully consistent with the Abel integral). As it
turns out, there are multiple oscillation frequencies in $F_{reg}(\omega)$
(see Appendix \ref{sec: Appendix-B}), and we cancel each of the four
dominant ones by a forth-order self-cancellation operation. Adapting
the notation of Appendix \ref{sec: Appendix-A} to the present specific
context (integration of $F_{reg}$), the desired generalized integral
is 
\begin{equation}
\int_{0}^{\infty(sc*)}F_{reg}(\omega)d\omega=\lim_{\omega\rightarrow\infty}H_{*}(\omega),\label{eq:  A34}
\end{equation}
where $H_{*}(\omega)\equiv T_{*}\left[H(\omega)\right]$, $H(\omega)$
is given in Eq. (\ref{eq:  A39}), and $T_{*}$ denotes the multiple
self-cancellation operation as generally defined in Eq. (\ref{eq:  A32})
and detailed in Eq. (\ref{eq:  A49}).

\begin{figure}
\begin{centering}
\includegraphics[bb=90bp 260bp 500bp 580bp,clip,scale=0.6]{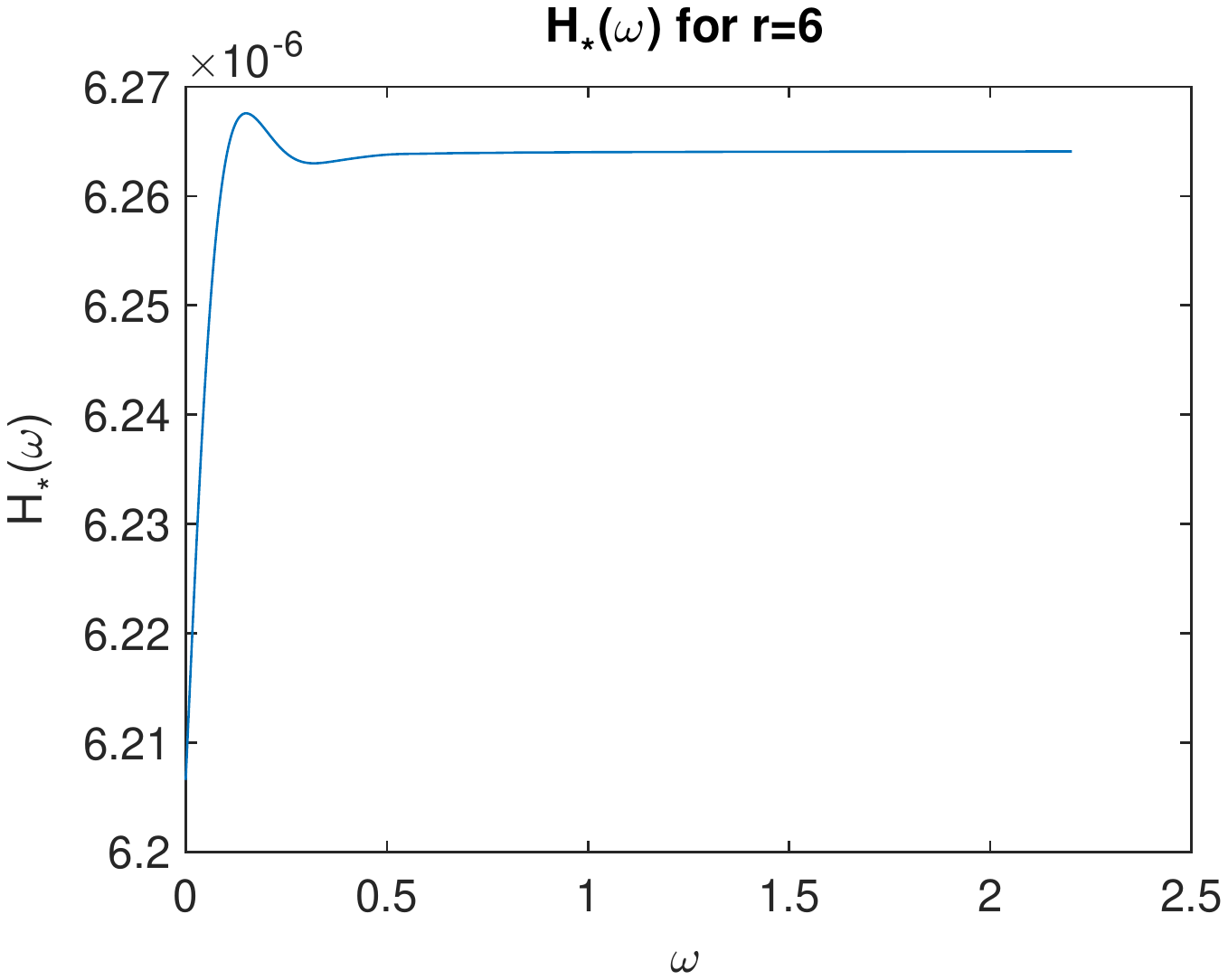}
\par\end{centering}

\protect\caption{The function $H_{*}(\omega,r=6)$ (obtained from $H$ after self-cancellation
of the oscillations). Notice the quick convergence. \label{fig:3}}
\end{figure}

\begin{figure}[H]
\subfloat[Comparing $\left\langle \phi^{2}\right\rangle _{ren}$ calculated
using our new method (the red pluses) to the previous calculation
by Anderson using the Euclidean sector and the WKB expansion.\label{fig:4a}]{\begin{centering}
\includegraphics[bb=70bp 260bp 500bp 590bp,clip,scale=0.47]{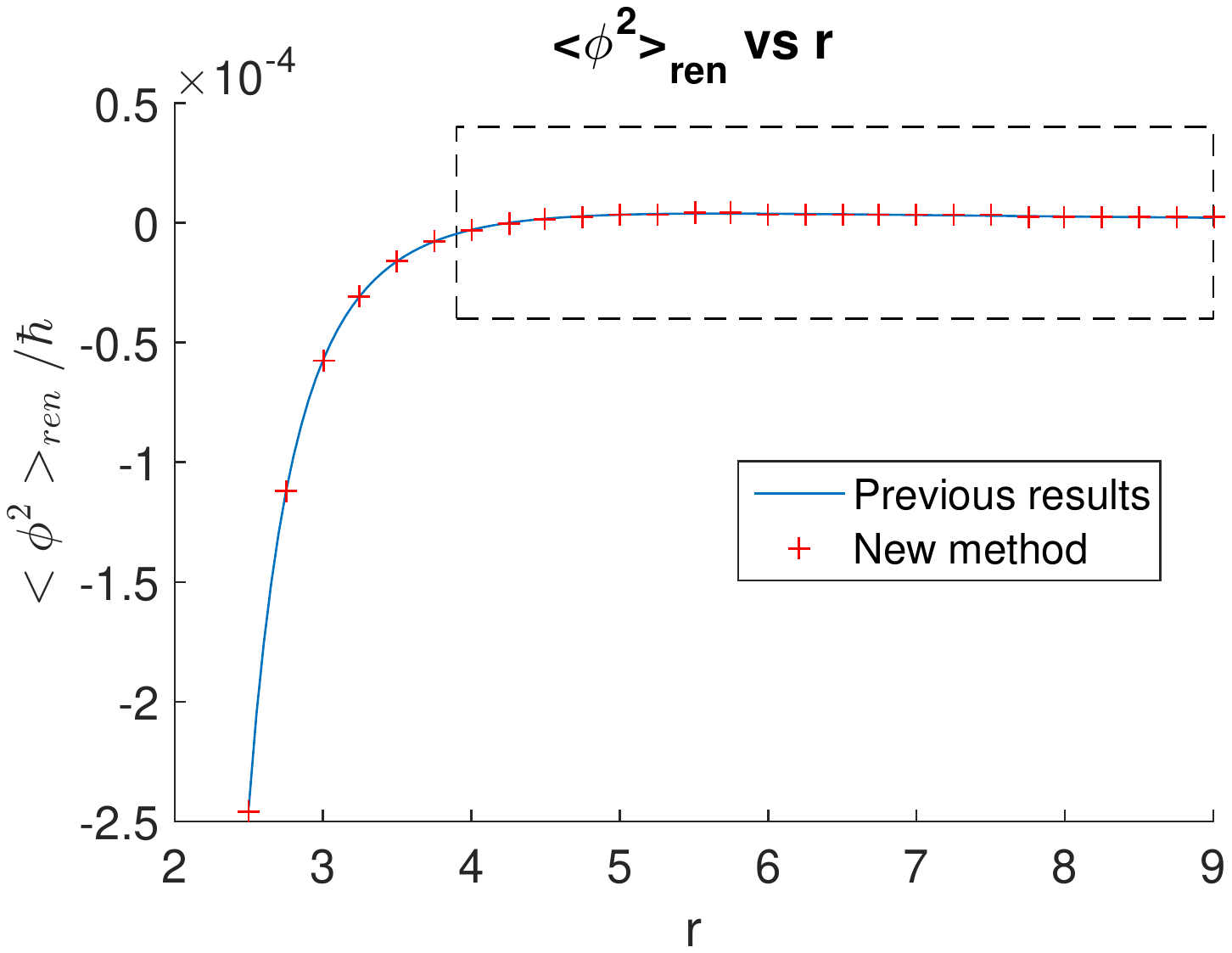}
\par\end{centering}

}\enskip{}\hfill{}\enskip{}\subfloat[Zoom-in on the box in Fig. \ref{fig:4a} for the values close to zero.\label{fig:4b}]{\begin{centering}
\includegraphics[bb=90bp 260bp 500bp 580bp,clip,scale=0.47]{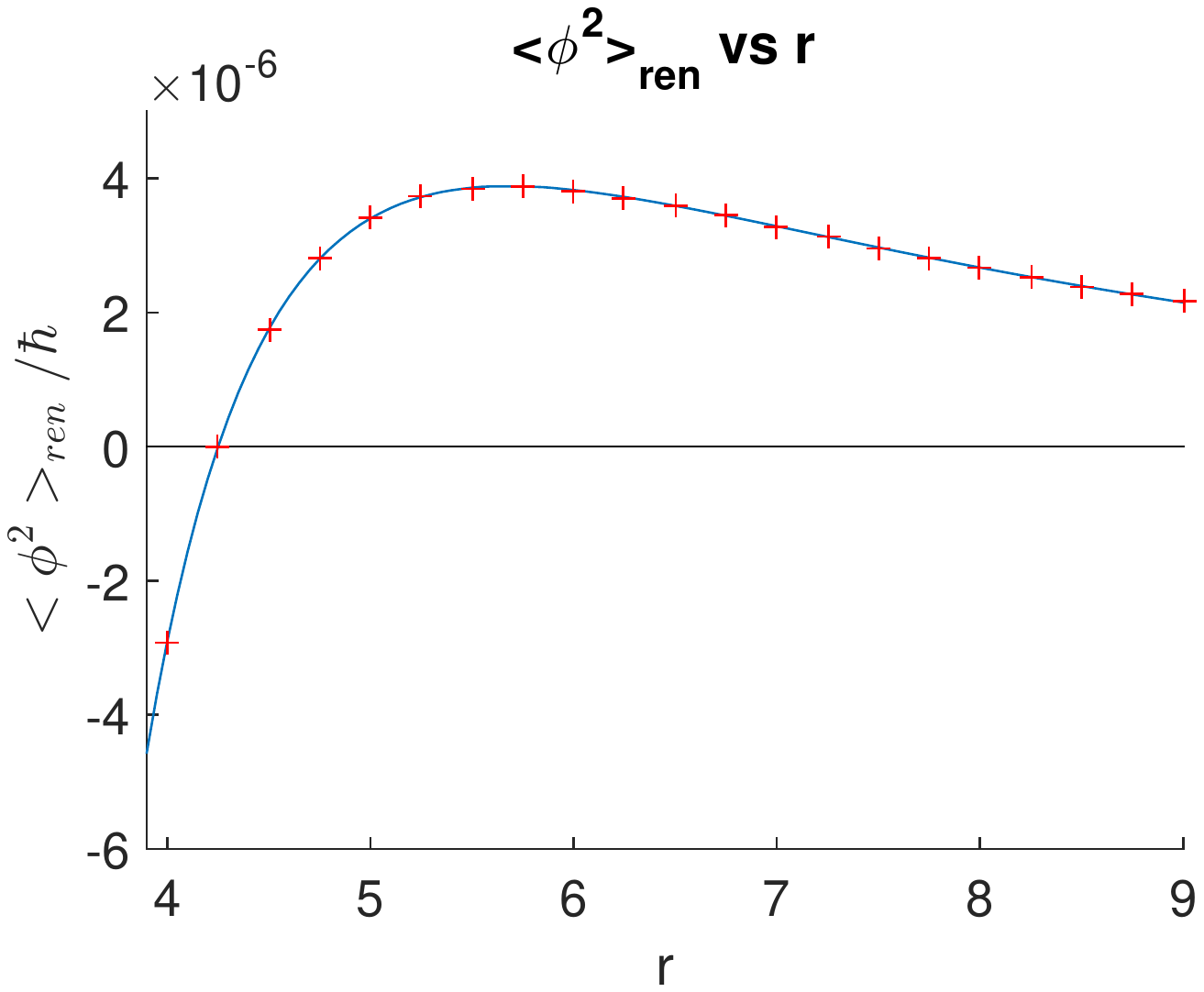}
\par\end{centering}

}

\protect\caption{}
\end{figure}

Figure \ref{fig:3} displays the function $H_{*}(\omega)$. It is
remarkable that after the oscillations have been removed the function
converges very quickly. This allows a fairly precise determination
of the limit $\omega\rightarrow\infty$ of this function, which constitutes
the generalized integral in Eq. (\ref{eq:  A34}). We then substitute
this integral, as well as $d(r)$ of Eq. (\ref{eq:  A10}), in Eq.
(\ref{eq:  A13}). In Figs. \ref{fig:4a} and \ref{fig:4b} we present
our results for $\left\langle \phi^{2}\right\rangle _{ren}$ as a
function of $r$, and compare them to results obtained previously
by Anderson \cite{Anderson (private)} using a very different method
(analytic extension to the Euclidean sector). The differences are
typically of order a few parts in $10^{3}$, consistent with the estimated
numerical errors.

\section{Discussion\label{sec: Discussion}}

We presented here a new approach for implementing point-splitting
regularization numerically, for the computation of $\left\langle \phi^{2}\right\rangle _{ren}$
and $\left\langle T_{\alpha\beta}\right\rangle _{ren}$ in various
asymptotically-flat spacetimes. Our main motivation in developing
this approach is to allow systematic investigation of self-consistent
semiclassical evaporation of BHs. This would require the calculation
of $\left\langle T_{\alpha\beta}\right\rangle _{ren}$ in a time-dependent
BH background.

So far we developed two variants of our basic approach (both for a
quantum scalar field in asymptotically-flat background): (i) $t$-splitting,
applicable to stationary spacetimes, and (ii) angular splitting, applicable
to spherically-symmetric spacetimes. In this first paper we focused
on the simplest of the two, the $t$-splitting variant. In presenting
this method, we restricted our attention to static spherically-symmetric
backgrounds, and to $\left\langle \phi^{2}\right\rangle _{ren}$ rather
than $\left\langle T_{\alpha\beta}\right\rangle _{ren}$, for the
sake of simplicity. But we also described the extension of the method
to more generic, non-spherical, stationary asymptotically-flat backgrounds.
This extension suffices for the non-BH case, e.g. for a rotating star.
In the case of a stationary BH which is not a spherically-symmetric
static one, our method still needs a completion: We still have to
formulate the construction of an appropriate ``up'' state at the
past horizon, which would constitute a physically-meaningful (and
properly Hadamard) ``vacuum'' state.

We then implemented the $t$-splitting variant to the specific case
of Boulware state in Schwarzschild spacetime (for a minimally-coupled
massless scalar field). The analytical part of the regularization
procedure coincides in this case with the one developed by Candelas
\cite{Candelas - 1979 - phi2 Schwarzschild}. However, here we also
implemented the numerical part, which involves the numerical solution
of the radial equation for the various modes, and the summation/integration
over the mode contributions. Doing so, we found that the regularized
function $F_{reg}\left(\omega,r\right)$ --- which was naively expected
to be well-behaved at large $\omega$ --- actually suffers from growing
oscillations, which make the $\omega$-integral non-convergent. As
it turns out, this phenomenon has little to do with the short-distance
behavior of the TPF in the coincidence limit $x'\rightarrow x$. Instead,
the oscillations originate from divergences of the TPF at \emph{remote}
points $x'$ which are connected to $x$ by null geodesics. We used
the notion of generalized integral --- and particularly the pragmatic
method of self-cancellation integral --- in order to handle these
oscillations and to carry the desired integration over $\omega$.
Doing so, we found excellent agreement with previous results obtained
by a different method, the Euclidean extension \cite{Anderson (private)}. 

Putting aside for a moment the ultimate goal of analyzing the time-dependent
evaporation process, we wish to emphasize that even the simplest version
of the method, the $t$-splitting variant presented here, makes it
possible to do PS regularization in non-spherical stationary spacetimes,
e.g. that of a strong-field spinning star (although in the generic
stationary BH case our construction still needs a completion). This
was not possible so far, due to the difficulties in conducting high-order
WKB expansion in such spacetimes.

However, to achieve our primary goal of self-consistent semiclassical
evaporation, we must deal with time-dependent backgrounds. To this
end we shall need the angular-splitting variant, which is slightly
more complicated than $t$-splitting. We shall describe this variant
in a separate paper. We already applied the angular-splitting method
to $\left\langle \phi^{2}\right\rangle _{ren}$ and also to $\left\langle T_{\alpha\beta}\right\rangle _{ren}$
in Boulware state in Schwarzschild, and again we found very good agreement
with previous calculations \cite{Anderson (private)}. These results
will be presented elsewhere \cite{Preparation}.

\section*{Acknowledgment}

We are grateful to Paul Anderson and Bob Wald for many interesting
and fruitful discussions. In addition AO thanks both Paul Anderson
and Bob Wald for kind hospitality in their institutions, where many
of these discussions took place. In particular the understanding of
the nature of the oscillations, and their relation to connecting null
geodesics, emerged in a discussion with Bob Wald during a visit at
the University of Chicago. We also wish to thank Paul Anderson for
sharing his unpublished data with us. 

\appendix

\section{Multiple self-cancellation\label{sec: Appendix-A}}

For a given integrand function $h(\omega)$, the self-cancellation
integral was defined in Sec. 2.2 as 
\begin{equation}
\int_{0}^{\infty(sc)}h(\omega)d\omega\equiv\lim_{\omega\rightarrow\infty}T_{\lambda}\left[H(\omega)\right],\label{eq:  A27-1}
\end{equation}
where
\[
H(\omega)\equiv\int_{0}^{\omega}h(x)dx,
\]
$T_{\lambda}$ is defined by 
\[
T_{\lambda}\left[f(\omega)\right]\equiv\frac{f(\omega)+f(\omega+\lambda/2)}{2},
\]
and $\lambda$ is the oscillation's wavelength. 

Consider next the situation in which $h(\omega)$ contains oscillations
in two different frequencies (say, with $\omega$-independent amplitudes).
Then $H(\omega)$ will take the asymptotic form 
\begin{equation}
H(\omega)\approx A_{1}e^{i\omega L_{1}}+A_{2}e^{i\omega L_{2}}+const\label{eq:  A51}
\end{equation}
with two different wave-lengths $\lambda_{1,2}=2\pi/L_{1,2}$. If
we apply a self-cancellation with respect to (say) $\lambda_{1}$,
then we will be left with a function $T_{\lambda_{1}}\left[H(\omega)\right]$
in which the $const$ is of course unaffected, and the term $\propto e^{i\omega L_{1}}$
has entirely been annihilated. The other oscillatory term $\propto e^{i\omega L_{2}}$
is still present in $T_{\lambda_{1}}\left[H(\omega)\right]$, although
its amplitude will decrease by the factor $\cos\left[(L_{2}/L_{1})\pi/2\right]$,
as one can easily verify. The self-cancellation integral (associated
to $T_{\lambda_{1}}$) is thus still non-convergent. A second self-cancellation,
this time with respect to $\lambda_{2}$, will yield a convergent
self-cancellation integral, which may be formulated as 
\[
\lim_{\omega\rightarrow\infty}T_{\lambda_{2}}T_{\lambda_{1}}\left[H(\omega)\right].
\]
The order of the two operations $T_{\lambda_{1}}$ and $T_{\lambda_{2}}$
is unimportant, as one can easily verify. With both orderings, the
double self-cancellation integral yields the ``$const$'' in Eq.
(\ref{eq:  A51}). %
\footnote{To avoid confusion, we point out that an application of self-cancellation
operation $T_{\lambda'}$, but with a ``mistaken'' wavelength parameter
$\lambda'$ that differs from the true oscillation wavelength $\lambda$
, does not ``spoil'' the generalized integral in any way: The mismatch
in $\lambda$ does not lead to any new oscillatory terms, nor does
it modify the value of the generalized integral. The effect of the
mismatch in $\lambda$ is merely to limit the efficiency of the self-cancellation
operation: It decreases the oscillation's amplitude by the factor
$\cos\left[\left(\lambda'/\lambda\right)\pi/2\right]$ instead of
fully annihilating it. %
}

This process is straightforwardly generalized to functions $H(\omega)$
with any number $k$ of oscillation frequencies that need to be annihilated.
To handle this situation, we define the multiple self-cancellation
operator 
\begin{equation}
T_{*}\equiv T_{\lambda_{1}}T_{\lambda_{2}}...T_{\lambda_{k}}.\label{eq:  A30}
\end{equation}
Then the multiple self-cancellation integral may be expressed as
\begin{equation}
\int_{0}^{\infty(sc*)}h(\omega)d\omega\equiv\lim_{\omega\rightarrow\infty}H_{*}(\omega),\label{eq:  A31}
\end{equation}
where 
\begin{equation}
H_{*}(\omega)\equiv T_{*}\left[H(\omega)\right].\label{eq:  A62}
\end{equation}

As was demonstrated in Sec. \ref{sub: Remarks-about-the-TPF}, if
$H(\omega)$ contains an oscillation with fixed amplitude, then the
$T_{\lambda}$ operation fully nullifies this oscillation. (This was
demonstrated there for the case $h(\omega)=\sin(\omega L)$, but it
equally applies to the more general cases $h(\omega)=e^{\pm i\omega L}$.)
However, if the oscillation's amplitude varies with $\omega$, the
cancellation is not complete. Consider the case $H(\omega)=g(\omega)e^{i\omega L}$
where $g(\omega)$ is some slowly-varying function --- namely, a function
whose typical length of variation $\ell$ becomes $\gg\lambda$ at
large $\omega$. To be more specific, let us further assume that $\ell$
diverges as $\omega\rightarrow\infty$. %
\footnote{This includes for example all powers $\omega^{p}$, because in this
class $\ell\propto\omega$. (But it also includes much more general
classes of functions, e.g. $\omega^{p}$$(\ln\omega)^{q}$ for any
$p,q$.)%
} Then one can easily show that 
\[
T_{\lambda}\left[H(\omega)\right]\simeq-\frac{\lambda}{2}\,\frac{dg}{d\omega}\, e^{i\omega L}
\]
at large $\omega$. In particular, if $H(\omega)=\omega^{p}e^{i\omega L}$
then 
\[
T_{\lambda}\left[H(\omega)\right]\propto p\,\omega^{p-1}e^{i\omega L}.
\]

Consider now the case $H(\omega)\approx\omega^{p}e^{i\omega L}$ for
some $1\leq p<2$. Then $T_{\lambda}\left[H(\omega)\right]$ still
does not converge as $\omega\rightarrow\infty$. Nevertheless, $T_{\lambda}\left(T_{\lambda}\left[H(\omega)\right]\right)$
is $\propto\omega^{p-2}e^{i\omega L}$ and hence it converges. This
illustrates that in certain circumstances one may want to apply the
same self-cancellation operation several times, say $n$ times, an
operation which we shall denote as $\left(T_{\lambda}\right)^{n}$. 

Quite generally, in the case $H(\omega)\approx\omega^{p}e^{i\omega L}$
one finds that %
\footnote{The exception is the case of natural $p$ with $n>p$, in which $\left(T_{\lambda}\right)^{n}[H(\omega)]$
strictly vanishes.%
} 
\[
\left(T_{\lambda}\right)^{n}[H(\omega)]\propto\omega^{p-n}e^{i\omega L}.
\]
Therefore, for this class of $H(\omega)$ functions, the convergence
criterion for the (single-frequency) multiple self-cancellation integral
is simple: It converges if and only if $n>p$. 

In the case $0<p<1$ a single self-cancellation operation would be
sufficient for achieving convergence. However, one would then be left
with a slowly-decaying amplitude $\propto\omega^{p-1}$. To speed
the large-$\omega$ convergence, one may repeat $T_{\lambda}$ several
times. This is in fact the situation in our specific problem, where
$h(\omega)\propto\omega^{1/2}e^{i\omega L}$ and hence $H(\omega)\propto\omega^{1/2}e^{i\omega L}+const$:
The single self-cancellation integral converges, but rather slowly,
with amplitude $\propto\omega^{-1/2}$, and we therefor repeat $T_{\lambda}$
several times to achieve faster convergence (which is crucial for
numerical implementation).

If there are several different oscillation frequencies that need to
be annihilated, one can freely choose how many times to repeat the
$T_{\lambda}$ operation for each frequency. We therefore generalize
the above expression (\ref{eq:  A30}) for the multiple self-cancellation
operator: 
\begin{equation}
T_{*}\equiv\left(T_{\lambda_{1}}\right)^{n_{1}}\left(T_{\lambda_{2}}\right)^{n_{2}}...\left(T_{\lambda_{k}}\right)^{n_{k}}.\label{eq:  A32}
\end{equation}
The multiple self-cancellation integral ($sc*$) is thus obtained
by using this $T_{*}$ operator in Eqs. (\ref{eq:  A31},\ref{eq:  A62}). 

Finally we point out that this multiple self-cancellation integral
is fully consistent with the Abel-summation integral --- in the same
sense discussed in Sec.\ref{sub: Remarks-about-the-TPF} (concerning
the single self-cancellation integral).

\section{The oscillations in $F_{reg}\left(\omega,r\right)$ \label{sec: Appendix-B}}

As can be clearly seen in Fig. \ref{fig:1b}, the function $F_{reg}\left(\omega,r\right)$
admits growing oscillations in $\omega$. These oscillations are of
course inherited directly from $F\left(\omega,r\right)$, as can be
seen in Fig. \ref{fig:1a} (although in the latter the oscillations
are overshadowed by the linearly-growing term). In this Appendix we
shall discuss the origin of these oscillations, their frequencies,
and their amplitudes. Then at the end we shall specify the self-cancellation
operator that we apply in order to practically remove these oscillations.

\subsection{Origin and Nature of oscillations}

Owing to spherical symmetry and staticity, in $t$-splitting the TPF
may depend only on $r$ and on $\varepsilon=t'-t$. Let us then introduce
the abbreviated notation for the TPF 
\[
P(\varepsilon,r)\equiv\left\langle \phi\left(t,r,\theta,\varphi\right)\phi\left(t+\varepsilon,r,\theta,\varphi\right)\right\rangle .
\]
Equation (\ref{eq: Our scheme - phi2 splitted}) actually tells us
that $F\left(\omega,r\right)$ is the Fourier transform of $P(\varepsilon,r)$.
\footnote{In this context we should regard $F\left(\omega,r\right)$ as a function
that vanishes for all $\omega<0$. Note also that in the present context
one should not think of $\varepsilon$ as a small parameter: Instead,
it is allowed to take all real values.%
} If $P(\varepsilon,r)$ were regular and smooth for all $\varepsilon$,
then its Fourier transform $F\left(\omega,r\right)$ would decay quickly
at large $\omega$, faster than any power of the latter. The undamped
oscillations in $F\left(\omega,r\right)$ must therefore indicate
some irregularity in $P(\varepsilon,r)$. We still need to understand
the nature of this singularity, and its location on the $\varepsilon$
axis. 

The singularity of the TPF at $\varepsilon\rightarrow0$ indeed leads
to a divergence of $F\left(\omega,r\right)$ at large $\omega$ (this
is the linearly-growing term shown in Fig. \ref{fig:1a}) --- but
not to oscillations; And this linear singularity is no longer present
in $F_{reg}\left(\omega,r\right)$. The oscillations in Fig. \ref{fig:1b}
must then indicate another singularity, located at some point $\varepsilon=\varepsilon_{s}\neq0$.
To illustrate this, consider for example the delta-function case:
The Fourier transform of $\delta(\varepsilon-\varepsilon_{s})$ is
$e^{-i\omega\varepsilon_{s}}$. It is oscillatory if and only if $\varepsilon_{s}\neq0$.
The frequency of the $\omega$-oscillations is just $\varepsilon_{s}$
--- namely, it directly tells us the distance of the singularity from
the point $\varepsilon=0$. 

One more example is the singularity $|\varepsilon-\varepsilon_{s}|^{-\beta}$,
whose transform is $\propto\omega^{\beta-1}e^{-i\omega\varepsilon_{s}}$.
But this phenomenon is of course more general: If a certain function
$P(\varepsilon)$ admits a Fourier transform $F\left(\omega\right)$,
then the transform of $P(\varepsilon-\varepsilon_{s})$ is $F\left(\omega\right)e^{-i\omega\varepsilon_{s}}$. 

The oscillations seen in e.g. Fig. \ref{fig:1b} have a certain ``$\omega$-wavelength''
$\lambda_{1}\approx0.17$ (in units in which the BH mass is $M=1$).
They must therefore correspond to a singularity in the TPF, located
at a distance $\varepsilon_{s}\equiv t'-t=2\pi/\lambda_{1}\approx37$
from point $x$. What is the nature of this non-local singularity
of the TPF? As already pointed out in Sec. \ref{sub: Remarks-about-the-TPF},
the function $\left\langle \phi\left(x\right)\phi\left(x'\right)\right\rangle $
admits a singularity whenever a null geodesic exists which connects
$x$ and $x'$. In the present context of Schwarzschild background
and $t$-splitting, we are dealing here with a null geodesic which
emanates from a certain spatial point $\left(r,\theta,\varphi\right)$,
makes a round trip around the BH, and then returns to that same spatial
point, but obviously with a certain delay in $t$ --- which should
correspond to the shift parameter $\varepsilon_{s}$.

\subsection{Spectrum of oscillations}

In fact there is not only one but an \emph{infinite,} discrete set
of such connecting null geodesics (for each $r$). This is because
a null geodesic emanating from a spatial point $\left(r,\theta,\varphi\right)$
can make any integer number of revolutions around the BH before returning
to that spatial point. Therefore, the TPF will actually admit an infinite
number of singular points at a discrete set of values $t'-t=\varepsilon_{n}$,
one for each integer $n$. 

Correspondingly, there will be a discrete spectrum of oscillation
modes in $F_{reg}\left(\omega,r\right)$, with ($r$-dependent) $\omega$-frequencies
$\varepsilon_{n}$ and corresponding wavelengths $\lambda_{n}=2\pi/\varepsilon_{n}$.
We point out, however, that the dominant mode is always $n=1$, and
the oscillation's amplitude quickly decays with $n$ (see next subsection). 

To perform the self-cancellation of oscillations, we shall need to
know the spectrum of frequencies $\varepsilon_{n}$, at any desired
$r$ value. This requires integration of the null geodesic equation,
to find the connecting null geodesics. For $r=3M$ the situation is
especially simple, because in that case the connecting null geodesic
is circular. One then finds that $\varepsilon_{n}=\left(2\pi\sqrt{27}\right)n$.
At other $r$ values the spectrum becomes more complicated, and needs
to be calculated numerically. The null orbits in Schwarzschild are
characterized by a single constant of motion, namely the angular momentum
per unit energy (or the ``impact parameter''). The calculation of
$\varepsilon_{n}$ involves (i) numerical integration of the null
geodesic equation (for prescribed values of that constant of motion),
and (ii) using the Newton-Raphson method for adjusting this constant,
in order to find the connecting geodesics, which return to the original
spatial point after $n$ rounds. The required parameters $\varepsilon_{n}$
are the $t$-duration of these connecting null geodesics. Overall,
this is an easy numerical procedure. 

In our representative case $r=6M$, the first few oscillation frequencies
are found to be 
\[
\varepsilon_{1}\simeq37.50\:,\;\;\;\varepsilon_{2}\simeq70.17\:,\;\;\;\varepsilon_{3}\simeq102.8\:,\;\;\;\varepsilon_{4}\simeq135.5\;.
\]
In Fig. \ref{fig:1b} we predominantly see the basic oscillation $n=1$,
and the frequency agrees very well with this value of $\varepsilon_{1}$.
We can also notice the residual effect of the $n=2$ oscillation,
which causes the small distortion (i.e. small deviation from the smooth
$\propto\omega^{1/2}$ envelope) in the pattern of peaks of $F_{reg}\left(\omega,r\right)$,
seen in Fig. \ref{fig:1b}. After the basic oscillation $n=1$ is
removed by self-cancellation, the next one ($n=2$) dominates and
becomes very clear, and again, its frequency is found to agree very
well with the above value of $\varepsilon_{2}$. By this procedure
it is possible to expose a few more modes, and to confirm their agreement
with the above $\varepsilon_{n}$ values, that were obtained from
the connecting null geodesics. 

As was already mentioned above, for $r=3M$ the frequencies $\varepsilon_{n}$
form an arithmetic sequence with a common difference $\Delta\varepsilon_{0}\equiv2\pi\sqrt{27}\simeq32.65$.
For other $r$ values this is no longer the case. Still, the difference
$\varepsilon_{n+1}-\varepsilon_{n}$ quickly approach the standard
spacing $\Delta\varepsilon_{0}$. For example, from the four $\varepsilon_{n}$
values specified above for the $r=6M$ case, one sees that $\varepsilon_{2}-\varepsilon_{1}\simeq32.67$,
and then $\varepsilon_{3}-\varepsilon_{2}\simeq\varepsilon_{4}-\varepsilon_{3}\simeq32.65$.
This is simply because for any $r$, a connecting null geodesic with
large $n$ makes most of the revolutions around the BH along an orbit
very close to the circle $r=3M$.

\subsection{Amplitude of oscillations}

The numerical data indicate that the oscillation's amplitude grows
as $\omega^{1/2}$. This in turn implies that the divergence of the
TPF at $\varepsilon\rightarrow\varepsilon_{s}$ should be $\propto|\varepsilon-\varepsilon_{s}|^{-3/2}$.
We shall not address this issue here in detail, but we point out that
this $-3/2$ power is just what one would expect from simple arguments.
To this end one has to recall that since the background is spherically
symmetric, and since the splitting is only in the $t$ direction (implying
that $\theta'=\theta$ and $\varphi'=\varphi$), whenever $x'$ is
located on a null geodesic emanating from $x$, it is placed \emph{exactly
}at a \emph{caustic point} of that null geodesic. Qualitative arguments
suggest that on crossing such a caustic point, the TPF should indeed
diverge as $|\varepsilon-\varepsilon_{s}|^{-3/2}$. But the discussion
of this issue is far beyond our present scope. 

The $n>1$ oscillations, too, grow as $\omega^{1/2}$ at large $\omega$
(the numerics confirms this, at least for the first few $n$ values).
This is for the same reason as that described above for $n=1$. We
thus express the large-$\omega$ amplitudes of the various $\varepsilon_{n}$
modes as $\approx A_{n}\omega^{1/2}$, where $A_{n}$ is a set of
($r$-dependent) amplitude parameters. 

It is important to explore how $A_{n}$ behaves with increasing $n$,
in order to control the possible effect of the infinite number of
oscillating modes. The simplest case to analyze is again $r=3M$,
because the connecting null geodesic is then the circular geodesic
at $r=3M$. Simple analytical considerations suggest that in this
case $A_{n}$ should form an almost-exact geometric sequence with
$A_{n}/A_{n+1}\cong e^{\pi}\simeq23.14$. %
\footnote{A simple (though still unproved) analytical argument, based on evaluating
the Van-Vleck determinant along the $r=3M$ geodesic, suggests that
in this case $A_{n}$ should be exactly proportional to $[2\sinh(2\pi n)]^{-1/2}$.
This expression deviates from the geometric sequence $e^{-\pi n}$
by a tiny relative amount $\cong e^{-4\pi n}/2$. This deviation is
smaller than one part in $10^{5}$ even for $n=1$, too small to be
detected by our numerics, but nevertheless our numerical results are
fully consistent with that expression.%
} Our numerical results for $F_{reg}\left(\omega,r=3M\right)$ allows
reliable evaluation of the first three amplitudes, and the calculated
ratios $A_{1}/A_{2}$ and $A_{2}/A_{3}$ agree very well with $e^{\pi}$,
to about one part in $10^{3}$. (For the $n>3$ modes the oscillations
are too weak to reliably measure their $A_{n}$.)

For other $r$ values the situation is more complicated, and we do
not expect to find such a well-approximated geometric sequence; Yet,
we still expect that as $n$ increases, $A_{n}/A_{n+1}$ should quickly
approach the above ``canonical'' value $e^{\pi}$. The reason is
that, for large $n$, the connecting null geodesic makes most of the
$n$ revolutions around the BH along an orbit very close to the circle
$r=3M$. Hence, the decrease in the Van-Vleck determinant at each
revolution is approximately the same as in the analogous $r=3M$ case
(an approximation that ever improves with increasing $n$). At very
small $n$, however, the ratio between two successive amplitudes may
slightly differ from $e^{\pi}$. We numerically find that the ratio
$A_{1}/A_{2}$ ranges from $e^{\pi}\simeq23.1$ at $r=3M$ to $\approx23.7$
at $r=9M$. 

Overall, at least in the range $3M\leq r\leq9M$ that we have numerically
explored, the numerical data as well as the theoretical considerations
are all consistent with an almost-geometric sequence (even for small
$n$), with $A_{n}/A_{n+1}$ ranging between $23$ and $24$. In turn
this implies that for practical computation of the generalized integral
of $F_{reg}\left(\omega,r\right)$, we shall have to cancel the first
few $n$ modes, but the contribution of large-$n$ modes may be neglected.

\subsection{Self-cancellation operator}

We self-cancel the first four frequencies $n=1...4$. The higher modes
$n>4$ are too weak to notice. %
\footnote{We point out that although the amplitude formally diverges at large
$\omega$ for \emph{any} $n$, because after all we only integrate
along a finite interval of $\omega$ {[}which is in turn allowed due
to the fast convergence of $H_{*}(\omega)${]}, and because of the
fast decay of $A_{n}$ with $n$, the large-$n$ oscillation terms
do not have any noticeable effect on the integral. %
} The values of the frequencies $\varepsilon_{n}$ are numerically
obtained from the connecting null geodesics, for any desired $r$,
as explained above. The corresponding wavelengths are then given by
$\lambda_{n}=2\pi/\varepsilon_{n}$. For each frequency, we apply
a fourth-order self-cancellation. Thus, in the terminology of Eq.
(\ref{eq:  A32}), our actual multiple self-cancellation operator
is 
\begin{equation}
T_{*}\equiv\left(T_{\lambda_{1}}\right)^{4}\left(T_{\lambda_{2}}\right)^{4}\left(T_{\lambda_{3}}\right)^{4}\left(T_{\lambda_{4}}\right)^{4}.\label{eq:  A49}
\end{equation}

\end{document}